\documentclass[a4paper,11pt]{article}
\pdfoutput=1 
\usepackage{jcappub} 
\usepackage{subcaption}
\usepackage{graphicx}
\usepackage{caption}
\usepackage{afterpage}

\title{\boldmath Photometric classification of type Ia supernovae in the SuperNova Legacy Survey with supervised learning}

\author[a,b,c,1]{A. M{\"o}ller,\note{Corresponding author.}}
\author[c]{V. Ruhlmann-Kleider,}
\author[c]{C. Leloup,}
\author[c,d]{J. Neveu,}
\author[c]{N. Palanque-Delabrouille,}
\author[c]{J. Rich,}
\author[e]{R. Carlberg,}
\author[f,b]{C. Lidman,}
\author[g]{C. Pritchet.}

\affiliation[a]{Research School of Astronomy and Astrophysics,\\ Australian National University, Canberra, ACT 2611, Australia.}
\affiliation[b]{ARC Centre of Excellence for All-sky Astrophysics (CAASTRO), Australia.}
\affiliation[c]{Irfu, SPP, CEA Saclay, \\ F-91191 Gif sur Yvette Cedex, France.}
\affiliation[d]{Universit{\'e} Paris-Sud, LAL UMR 8607, F-91898 Orsay Cedex, France \& CNRS/IN2P3, F-91405 Orsay, France.}
\affiliation[e]{Department of Astronomy and Astrophysics, University of Toronto,\\ 50 St. George Street, Toronto, ON M5S 3H8, Canada.}
\affiliation[f]{Australian Astronomical Observatory,\\ North Ryde, NSW 2113, Australia.}
\affiliation[g]{Department of Physics and Astronomy, University of Victoria,\\ P.O. Box 3055, Victoria, BC V8W 3P6, Canada.}

\emailAdd{anais.moller@anu.edu.au}
\emailAdd{vanina.ruhlmann-kleider@cea.fr}
\emailAdd{clement.leloup@cea.fr}
\emailAdd{jneveu@lal.in2p3.fr}
\emailAdd{nathalie.palanque-delabrouille@cea.fr}
\emailAdd{james.rich@cea.fr}
\emailAdd{raymond.carlberg@utoronto.ca}
\emailAdd{chris.lidman@aao.gov.au}
\emailAdd{pritchet@uvic.ca}

\abstract{In the era of large astronomical surveys, photometric classification of supernovae (SNe) has become an important research field due to limited spectroscopic resources for candidate follow-up and classification. In this work, we present a method to photometrically classify type Ia supernovae based on machine learning with redshifts that are derived from the SN light-curves. This method is implemented on real data from the SNLS deferred pipeline, a purely photometric pipeline that identifies SNe Ia at high-redshifts ($0.2<z<1.1$). 

Our method consists of two stages: feature extraction (obtaining the SN redshift from photometry and estimating light-curve shape parameters) and machine learning classification. We study the performance of different algorithms such as Random Forest and Boosted Decision Trees. We evaluate the performance using SN simulations and real data from the first 3 years of the Supernova Legacy Survey (SNLS), which contains large spectroscopically and photometrically classified type Ia samples. Using the Area Under the Curve (AUC) metric, where perfect classification is given by 1, we find that our best-performing classifier (Extreme Gradient Boosting Decision Tree) has an AUC of $0.98$.

We show that it is possible to obtain a large photometrically selected type Ia SN sample with an estimated contamination of less than $5\%$. When applied to data from the first three years of SNLS, we obtain 529 events. We investigate the differences between classifying simulated SNe, and real SN survey data. In particular, we find that applying a thorough set of selection cuts to the SN sample is essential for good classification. This work demonstrates for the first time the feasibility of machine learning classification in a high-$z$ SN survey with application to real SN data.
}

\begin{document}
\maketitle
\flushbottom

\section{Introduction}

Type Ia supernovae are used as standard candles to measure the expansion history of the Universe. Since the discovery of the accelerated expansion of the Universe \citep{Riess:1998uy,Perlmutter:1999tu}, enormous effort has been dedicated to obtaining larger samples of SNe Ia with high quality light-curves. Second-generation surveys such as SNLS and SDSS-II \citep{Betoule:2014ui} have obtained such samples using spectroscopic classification of the SN type. However, in the era of large surveys such as DES \citep{Bernstein:2011zf} and LSST \citep{Abell:2009aa}, spectroscopic resources are insufficient for complete candidate follow-up and classification, making purely photometric classifications necessary. 

The SNLS deferred photometric pipeline can be separated in two parts: i) the detection of SN-like events, and ii) their classification as type Ia SNe \cite{Bazin:2011em} or core-collapse SNe \cite{Bazin:2009mp}. In the analysis of the 3-year SNLS data (SNLS3), this pipeline provided a sample of $486$ photometrically identified SNe Ia, almost twice the number of spectroscopically identified type Ia SNe found by the SNLS real-time analysis pipeline. Classification in the deferred pipeline included the use of host-galaxy photometric redshifts. Redshifts were assigned by matching SNe to host-galaxies with photometric redshifts in the \textit{Ilbert} catalog \citep{Ilbert}. This assignment had an efficiency of $83\%$. Events without a photometric redshift could not be classified in that analysis and were subsequently removed.

This paper presents a new photometric classification of SNLS SNe based on supervised learning, with redshifts derived directly from SN light-curves. The redshift algorithm was trained on SNLS3 data and has better average precision and fewer catastrophic errors than the host galaxy photometric redshift catalog used in the previous analysis. It provides redshifts for all SNe and is independent of cosmological parameters. The redshift algorithm is described in more detail in \citep{PalanqueDelabrouille:2009ng}.

In order to exploit all the available information and to optimize classification, we take advantage of machine learning algorithms, notably Boosted Decision Trees (BDTs). BDTs are supervised learning methods where the algorithm learns from a known ``training sample" before classifying unidentified data. Supervised learning methods have been previously used for photometric SN classification, e.g.  the SuperNova Photometric Classification Challenge \citep{Kessler:2010wk}, with good results \citep{Kessler:2010qj,Karpenka:2012pm,Ishida:2012cf,Lochner:2016hbn}. In this work, the performance of the supervised learning classification is estimated not only with simulated SNe but also with the large type Ia photometric sample obtained in the previous SNLS3 analysis \citep{Bazin:2011em}. This work therefore sets a precedent for the application of machine learning methods to the classification of real SN data.

The outline of the paper is as follows: the SNLS data and simulations are presented in Section 2. In Section 3, we describe the measurement of light-curve parameters as well as the selection of light-curves of sufficient quality to be treated for classification. In Section 4, we introduce the classification algorithms used in this work and the metrics used for evaluation. We compare different classification algorithms in Section 5 using simulated, photometric and spectroscopic SN samples. Our best performing method is studied in detail in Section 6. We summarize and conclude this work in Section 7.

\section{SNLS data and SN simulation}

SNLS is part of the Deep Synoptic Survey conducted at the Canada-France-Hawaii Telescope (CFHT). Using a rolling-search strategy, it targeted four one square degree fields during 5 to 7 consecutive lunations per year over a period spanning five years using four different broadband filters $g_M$, $r_M$, $i_M$ and $z_M$ \citep{Regnault2009}. Two independent analysis pipelines processed SNLS data. The first, which we refer to as the real-time pipeline, relies on the spectroscopic follow-up of detected SN candidates for classification and redshift determination \citep{Guy:2010bp}. The second is the deferred photometric pipeline.

This work is based on the deferred photometric pipeline. It is independent of the real-time analysis and requires only photometric information. In this pipeline, transient events are detected in one filter and multi-band light-curves are processed for all detections. Then, a set of cuts described in \cite{Bazin:2011em} is applied to reject spurious objects and obtain a sample of events whose light-curves are consistent in shape with that expected from SNe, hereafter referred as SN-like events. Classification of these SN-like events is the subject of this work. We will classify SNe into two types: type Ia SNe with correct redshifts (``the signal") and ``the background" which consists of other types of SNe or type Ia SNe with inaccurate redshifts. In this work, redshifts are considered accurate if, when compared to the generated (spectroscopic) redshift for simulations (data), they satisfy $|\Delta z|/(1+z) < 0.1$, inaccurate redshifts exceed $0.15$. More details on the deferred photometric pipeline can be found in \cite{Bazin:2011em}.

\subsection{SN simulation}
To set up the classification procedure, we use synthetic type Ia and core-collapse SNe that were generated for the SNLS3 analysis in \cite{Bazin:2011em}. This allows for a more direct comparison between this work and the method used in \cite{Bazin:2011em}. 

Synthetic light-curves of SNe Ia within the redshift range $(0,1.2]$ were produced in \cite{Bazin:2011em} with SALT2 \citep{Guy:2007dv} assuming a flat Universe with $\Omega_m =0.23$. Simulated SNe Ia were generated assuming a constant co-moving volumetric rate. Values of the SALT2 color and $X_1$ parameters ($X_1$ is related to the light-curve width) were randomly selected from Gaussian distributions that match the distributions of the spectroscopically identified SNe Ia. To represent the SNIa population before selection effects, only SNe Ia with $z<0.7$ were used to parametrize the distributions \citep{Perret2010}. For each synthetic SN, a random position in the SNLS fields was assigned, Milky Way dust extinction corrections were applied and detection effects were simulated. From the original list of $20,000$ simulated SNe Ia, $49\%$ were detected and passed the SN-like selection cuts. The majority of the lost supernovae had low signal-to-noise ratios.

To increase the number of events, we chose not to impose the detection criteria in \cite{Bazin:2011em}. This allows a larger number of faint events to be used in the classification. We keep selection cuts for consistency. In total $10,522$ SNIa synthetic light-curves are used in this work.

Core-collapse supernovae (CC SNe) can be separated into those that have a plateau in the light-curve after maximum light, and those that do not. Synthetic light-curves were generated for both CC types in the redshift range $0<z<1.2$ assuming a constant co-moving volumetric rate. A light-curve model \cite{Bazin:2011em} was constructed based on the sample of 117 SNLS CC SNe at $z<0.4$ \citep{Bazin:2009mp}. From 40,000 synthetic light curves, 20,000 of each type, only $10.5\%$ were detected and selected as SN-like. The efficiency is very low because of the low luminosity of CC SN compared to SNIa. As for synthetic type Ia SNe, for our classification we chose not to impose the detection criteria used in \cite{Bazin:2011em} and obtain a sample of $~5000$ simulated core-collapse SNe.

\subsection{SNLS3 data}\label{section:SNLS3_data}
\begin{figure}[h!]
\centering
\begin{center}
\begin{subfigure}[b]{0.48\textwidth}
\includegraphics[width=0.8\linewidth]{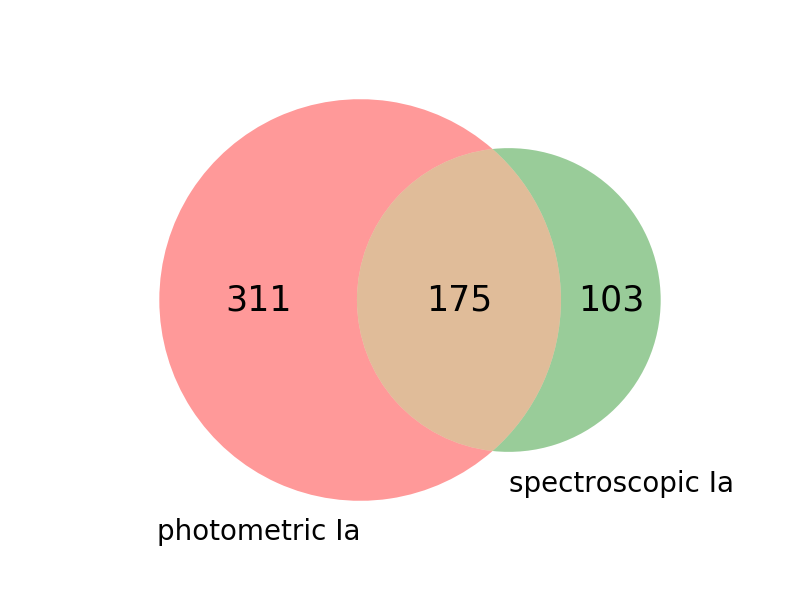}
\caption{SNe Ia: 175 common events.}\label{fig:venn_Ia_spe_photo}
\end{subfigure}
\begin{subfigure}[b]{0.48\textwidth}
\includegraphics[width=0.8\linewidth]{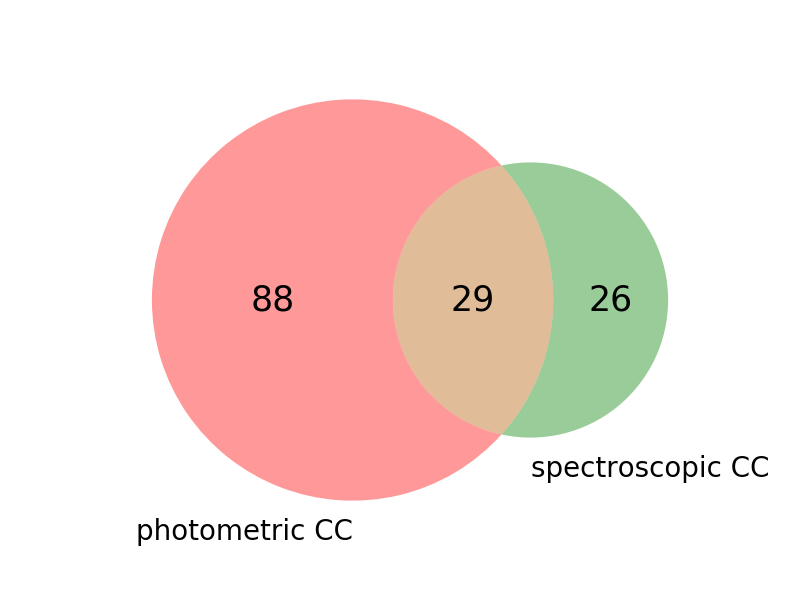}
\caption{CC SNe: 29 common events.}\label{fig:venn_CC_spe_photo}
\end{subfigure}
\caption{Venn diagrams of spectroscopically and photometrically identified type Ia (\ref{fig:venn_Ia_spe_photo}) and CC (\ref{fig:venn_CC_spe_photo}) from the SNLS3 classification. Supernovae common to both samples are shown in beige.}\label{fig:venn_spe_photo}
\end{center}
\end{figure}

To evaluate the performance of our classification, we use not only simulated events but also SNe that have been identified from the first 3 years of SNLS (hereafter referred to as SNLS3). These SNe were identified photometrically or spectroscopically by independent analyses. The photometric and spectroscopic samples contain a number of events in common as illustrated in Figure \ref{fig:venn_spe_photo}.

As previously mentioned, in the SNLS deferred photometric pipeline, classification is performed on events that are first detected as transients and then selected as SN-like. This SN-like sample is the starting point of our classification and consists of $1483$ events.

\subsubsection{Type Ia SNe}
The spectroscopically and photometrically identified type Ia samples are shown in Table \ref{table:SNLS3}. Events common to both samples are illustrated in Figure \ref{fig:venn_Ia_spe_photo}. In SNLS3, $486$ events were photometrically classified as SNe Ia with an estimated purity of $94.4\pm 0.5\%$ \citep{Bazin:2011em}. 

In this work, we consider two sub-classes of spectroscopic SNe Ia, split according to the confidence index (CI) of the spectroscopic identification: certain SNIa are denoted ``SNIa" (corresponding to $CI=4$ and $CI=5$ in the classification scheme of \citep{Howell}) and probable SNIa ($CI=3$) are labeled ``SNIa*". Further details about these indices can be found in \citep{Howell}.

A photometrically classified sample of 18 subluminous SNe Ia at $z<0.6$ was obtained in \citep{GonzalezGaitan:2010iw}. The SN-like sample contains 16 of these 18 subluminous SNe Ia. SNLS also detected 8 SNe Ia that were spectroscopically classified as peculiar. This sample includes super-Chandrasekhar and 1991T-like SN events \citep{Balland:2009ka,Bronder:2007hp,Ellis:2007hx,Howell:2006vn}.

\subsubsection{Core-collapse SNe}
In the SNLS3 SN-like sample, $55$ events were identified spectroscopically as core-collapse SNe \citep{Bazin:2011em}. A photometric classification based on the deferred pipeline identified $117$ events as core-collapse SNe with an estimated purity of  $97\%$ \citep{Bazin:2009mp}. Common events between both samples are illustrated in Figure \ref{fig:venn_CC_spe_photo}.

\begin{table}
\centering
\begin{tabular}{|c | c |c | c |}
	\hline
	SNLS3& \multicolumn{3}{|c|}{SNe Ia} \\
	 & \multicolumn{2}{|c|}{spectroscopic} & photometric \\
	 \hline
	 SN-like & \multicolumn{2}{|c|}{278} & 486 \\
	 & \multicolumn{2}{|c|}{ } & \\
	 CI & SNIa & SNIa* &\\
	 & 213 & 65 &  \\
	 \hline
\end{tabular}
\caption{Type Ia SNe in SNLS3. The number of events identified spectroscopically (photometrically) in the SN-like sample are shown. For spectroscopic SNe Ia, we further divide the sample using the confidence indices (CI) as defined in \citep{Howell}. }
\label{table:SNLS3}
\end{table}

\section{Light-curve analysis before classification}

\begin{figure}[h!]
\begin{center}
\includegraphics[width=0.7\linewidth]{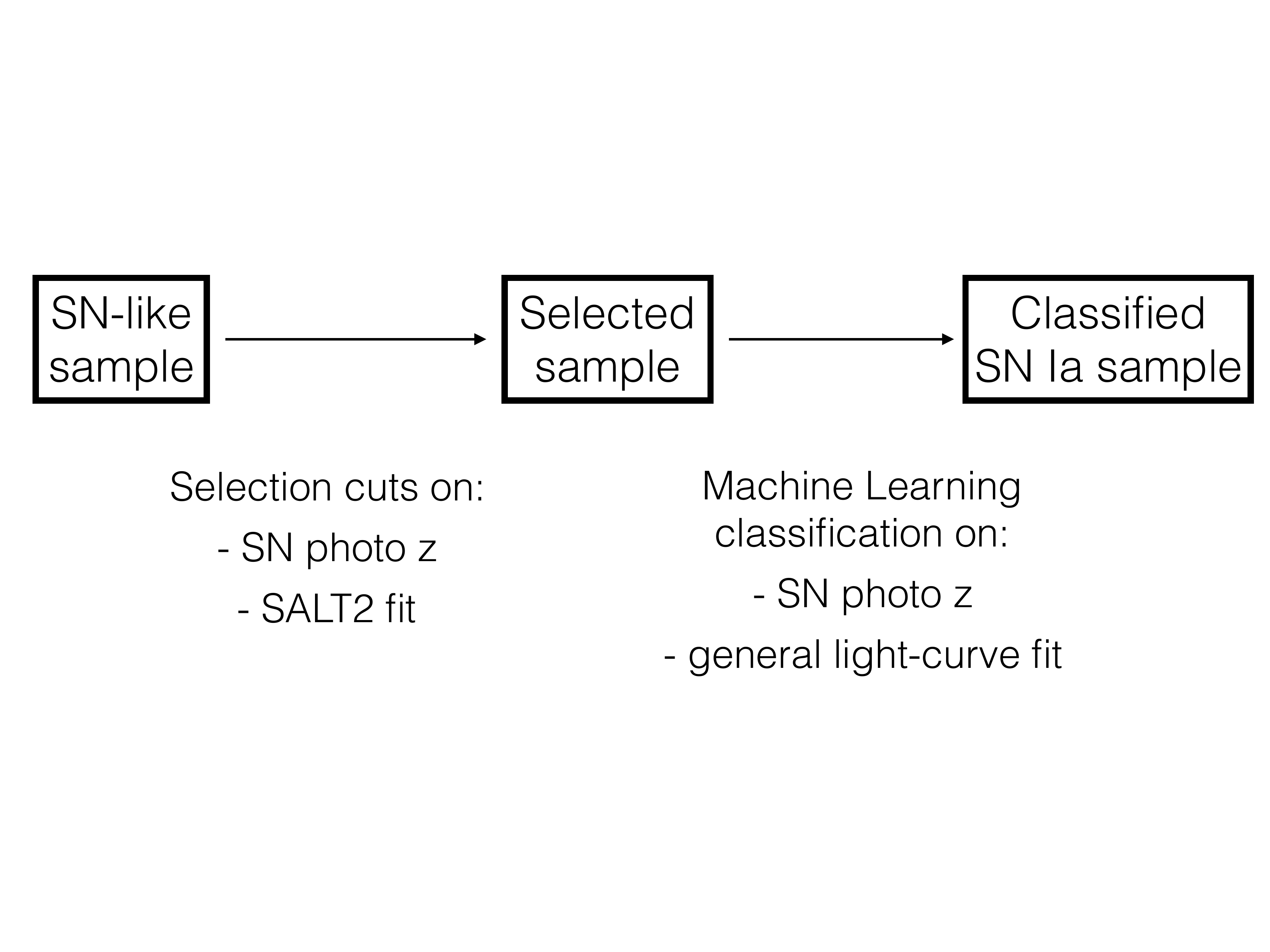}
\caption{Flow chart showing the selection and classification steps used in this work. The SN samples are shown as solid lines. $SN photo z$ stands for photometric SN redshifts.}\label{fig:flow_chart}
\end{center}
\end{figure}

Before classification, all SN-like events are processed. First, photometric SN redshifts are obtained by using the algorithm described in Section \ref{Section:SN_photo_z}. Then, selection cuts are applied on the photometric SN redshift quality and the light-curve quality as determined by a SALT2 fit. Finally, events are fitted with a general light-curve fitter. Only the photometric SN redshift and the general light-curve fitter parameters are used in the classification. The complete procedure is illustrated in Figure \ref{fig:flow_chart}. In the following we introduce the algorithms and selection cuts used in our analysis.

\subsection{Photometric SN redshifts}\label{Section:SN_photo_z}

We use the algorithm elaborated in {\it Palanque-Delabrouille et al.} \citep{PalanqueDelabrouille:2009ng}, trained with SNLS3 data, to obtain redshifts for each SN-like event. These redshifts will be hereafter called photometric SN redshifts. The algorithm obtains the redshift of type Ia SNe using the SALT2 light-curve fitter in a iterative procedure. In the first iteration, successive values of redshift are fit while color and stretch SALT2 parameters are constrained by priors. When a solution if found, another scan is done around the fitted redshift with free color and stretch parameters. 

The precision of the redshifts for SNe in the SNLS3 sample was reported in \citep{PalanqueDelabrouille:2009ng}. The average precision was defined as $\sigma_{\Delta z/(1+z)}\equiv 1.48 \times$ median $[|\Delta z| / (1+z)]$, where $\Delta z$ is the difference between the real and the photometric SN redshift. The rate of catastrophic errors, $\eta$, was defined as the proportion of events with $|\Delta z|/(1+z) > 0.15$. For the SNLS3 sample an average precision of $\sigma_{\Delta z/ 1+z}=0.022$ up to $z\sim1$ was found, while for $z < 0.45$ it was $0.006$. The precision degrades with redshift due to low flux, first in the $g$ band and then in the $r$ band as redshift increases. This degradation is irregular as seen in Fig. 4 and described in Section 5 in \citep{PalanqueDelabrouille:2009ng}. Catastrophic errors were found to be under $1.4\%$ for type Ia SNe passing color and stretch cuts. When restricting the test sample to spectroscopically confirmed SNe Ia the catastrophic errors fell to $0.4\%$. The authors found a net bias on the fitted redshift of 0.008 on average. Further details can be found in \citep{PalanqueDelabrouille:2009ng}.

\begin{figure}[h!]
\begin{center}
	\includegraphics[width=.7\linewidth]{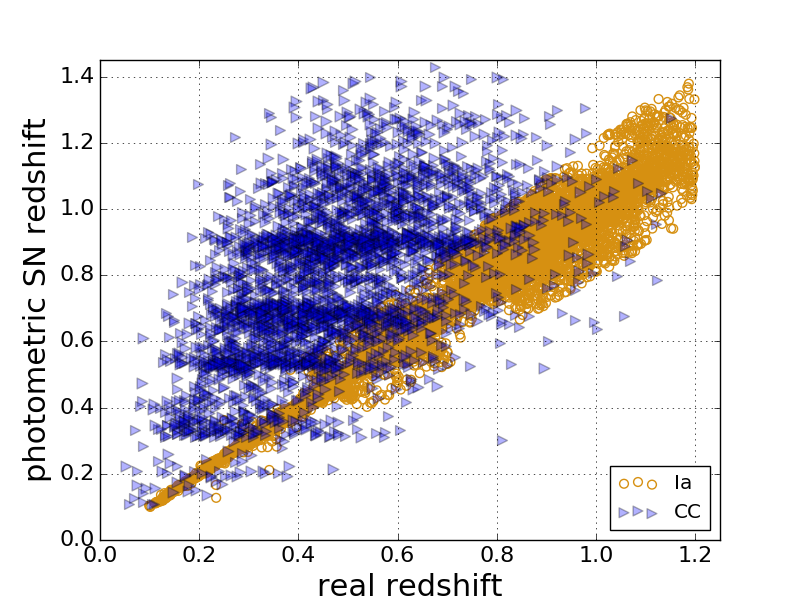}
	\caption{Photometric SN redshift vs. the real redshift for simulated SNe. We plot type Ia SNe with yellow circles and core-collapse SNe with blue triangles.}\label{fig:sim_SN_zpho_gz}
\end{center}
\end{figure}

The algorithm can be used for obtaining redshifts for all SN-like events. In Figure \ref{fig:sim_SN_zpho_gz}, we plot the light-curve redshift against the real redshift for simulated type Ia and core collapse SNe. Since the algorithm assumes that all events are SNIa, redshifts obtained for core-collapse SNe are usually inaccurate. However, there are some core-collapse SNe that have redshifts that are close to correct. These events have colors that are consistent with type Ia SNe.

The photometric SN redshift algorithm provides not only the redshift, but also the reduced $\chi_\nu^2$ of the agreement in color and width between the processed light-curve and the expected light-curve of an SNIa at the determined redshift. In the following we will refer to the reduced chi-square, which is defined as $\chi^2_\nu = \chi^2/N_{dof}$, there $N_{dof}$ is the number of degrees of freedom.

\subsection{Selection cuts}\label{Section:selectioncuts}
Before classification, we apply selection cuts to ensure meaningful photometric redshifts and reliable light-curves, and to mitigate the number of non-SN events still present in the SN-like sample. Indeed, since the identification of SN-like events in the photometric pipeline was designed to be sensitive to different types of SNe and to faint ones, non-SNe are probably still present and may bias the classification. We thus restrict the sample further by applying cuts more focused on type Ia events. 

\subsubsection{Photometric SN redshift quality}
We assess the quality of photometric SN redshifts through the goodness of the light-curve fit. The photometric SN redshifts algorithm performs iterative fits and in each passing total $\chi^2$ and contributions to the total $\chi^2$ by the priors can be obtained. More details can be found in \citep{PalanqueDelabrouille:2009ng}.

\begin{figure}[h!]
\begin{center}
\includegraphics[width=0.8\linewidth]{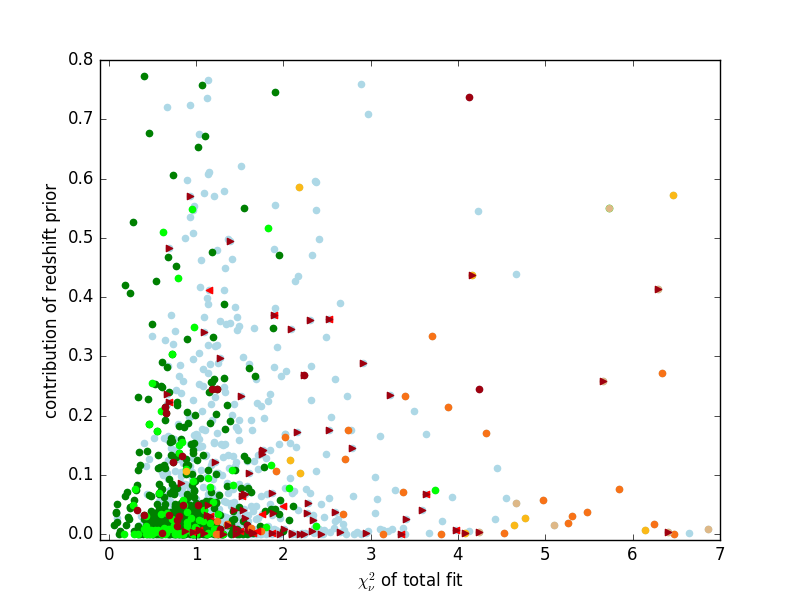}
\caption{The contribution of the redshift prior to the total reduced $\chi^2_\nu$ for SNLS3 SN-like events. Events classified as type Ia are in light (dark) green for spectroscopically confirmed (photometrically identified) supernovae. Core-collapse SNe are in bright red left-pointing (dark red right-pointing) triangles for spectroscopic (photometric) sample. Orange (yellow) dots stand for events whose light-curves were visually inspected and found not to be compatible with a SN-like signal (compatible with long declining events similar to SNe II). Beige dots are events with incompatible photometric and spectroscopic redshifts. All other SN-like events of unknown type are in blue. SNe with a total $\chi^2_\nu$ that is greater than four are excluded from our selection.}\label{fig:non-physical}
\end{center}
\end{figure}

To investigate possible cuts, we visually inspected a subset of the light curves that are outliers in a number of diagnostic plots, such as the plot shown in Fig. \ref{fig:non-physical}. In particular, the $\chi^2_\nu$ of the total multi-band fit was found to be sensitive to non SN-like events. We chose to exclude events with a total $\chi^2_\nu$ greater than four. 

Cuts were also derived for other output variables of the photometric SN redshift algorithm such as the $\chi^2_\nu$ of the redshift, color and stretch. 

\subsubsection{Light-curve quality}
With the redshift fixed to the photometric redshift we now assess the quality of our light-curves. For this, we fit our light-curves with SALT2. The output of this fit is used only to remove events with insufficient light-curve coverage and events with poor fits. It provides a fitted date of maximum light as well as color, stretch and peak magnitudes. It is not used in the classification since SALT2 was already used for obtaining photometric SN redshifts. We require:

\begin{itemize}
\item Minimal sampling of the light-curve before and after the SALT2 fitted time of maximum in rest frame, $\tau$:
	\begin{itemize}
	\item at least one measurement in the range $-10<\tau<+5$ days,
	\item at least one measurement in the range $+5<\tau<+20$ days for a reasonable shape evaluation,
	\item at least one measurement in each band from a pair selected from $(g-i)$, $(r-z)$ or $(i-z)$ must be within the range $-10<\tau<+35$.
	\end{itemize}
\item SALT2 convergence: events for which the SALT2 fit did not converge are discarded.
\end{itemize} 

\subsection{Light-curve fitter}\label{Section:feature_extraction}
To parameterize the light-curve shape we use the functional form \citep{Bazin:2011em}:
\begin{equation}
f^{k} (t)=A^k \frac{\exp{-(t-t_0^k)/\tau^k_{fall}}}{1+\exp{-(t-t_0^k)/\tau^k_{rise}}} + c^k \; ,
\label{eq:sn-like_var}
\end{equation}
where $A^k$ sets the normalization, $\tau^k_{fall}$ ($\tau^k_{rise}$) defines the fall (rise) time, $t_0^k$ is related to the date of maximum as $t^k_{max}= t^k_0 + \tau^k_{rise} \ln (\tau^k_{fall}/\tau^k_{rise} -1)$ and $c^k$ is a constant. 

First, the flux in each filter, $k$, is fitted. Then, we impose that all fits share the same $t_0$ from the $i$ band \footnote{Some SNe have different maximum dates for different filters. This requirement was set for consistency between fits. It is not expected that this affects the classification.}. A second fit is done using Gaussian priors from the first fit.

The fitting procedure provides the amplitude, rise and fall times for each filter. These are relevant features to characterize a SN light-curve and are the ones used for the following SN classification.

\section{Classification with Machine Learning}

Our goal is to select type Ia SNe from a SN sample. This can be reduced to a problem of predicting the type for each event. Machine learning algorithms provide an automated way of classifying events. In particular, supervised learning algorithms can learn from data in order to make predictions. 

The features available for the classification are: the photometric SN redshift, the color and stretch obtained from the redshift fit and their $\chi^2_\nu$, the values for the general light-curve fit in each band (amplitudes, rising and falling times) and the $\chi^2_\nu$ of the general light-curve fits.

In this section, we will first introduce the machine learning algorithms used in this work.\footnote{We rely on the Python package scikit-learn \citep{scikit-learn} for implementation.} Then, we will introduce our model validation technique, cross-validation. Performance is evaluated using metrics that will be described in Section \ref{section:metrics}.

\subsection{Boosted Decision Trees (BDTs)}
BDTs are supervised classification methods that perform well with large data sets and are adapted to classify high-dimensional data. From the \textit {training sample} they learn a mapping function that allows them to classify other data points (\textit {classification sample}). Their output is a \textit{prediction}: the probability of an object to belong to a given class (BDT response).

A decision tree (DT) makes successive rectangular cuts in the parameter space to classify data. Binary splits separate the data into subsamples (``leaf nodes") which at the end of the tree are given a probability to be classified as signal or background (prediction). At each split, the algorithm determines the variable that gives the best separation to discriminate between signal and background (in terms of classification error). Often, trees are too complex and do not generalize to other samples (they ``overfit"). To avoid this, trees can be combined to improve generalizability and stability. Two main approaches for combining trees are averaging and boosting methods.

Averaging methods construct several estimators (decision trees). The final prediction is an average of all the DTs prediction. Such methods are:
\begin{itemize}
\item Random Forest (RF): $n$ decision trees are constructed from a sample drawn with replacement (an event can be drawn multiple times) from the training set. In the learning process, the feature to be used at each binary split is picked from a random subset of the features (which can include or not the best available feature). This is done to obtain a better model where variance is decreased. The final prediction contains an average of all probabilistic predictions in different trees.

\item Bagging: decision trees are built from random subsets of the training set. For each event, the final prediction is the sum of the predictions from all trees.
\end{itemize}

Boosting methods build a model iteratively. They combine ``weak" classifiers as small decision trees on modified versions of the training data. We will use:
\begin{itemize}
\item AdaBoost (AB): iteratively constructs an additive model for the data. At each iteration, the data set is classified and each individual event is given a weight which represents its importance in the classification. The weight of individual events on the training sample is modified at each iteration. Those events that were incorrectly predicted at the previous step have their weights increased and those that were correctly predicted have their weights decreased. In this sense, misclassified events are the focus of the next iteration. All predictions are combined by a weighted sum to produce the final prediction. 

\item XGBoost (XGB): constructs an additive model while optimizing a loss function. The loss function accounts for the inaccuracy of predictions in the classification. The performance is given by an objective function that contains both a loss function and a regularization term (controls complexity of the model). This is a more refined version of GradientBoosting which is accurate and has shown good performance in classification challenges \citep{xgb} \footnote{http://xgboost.readthedocs.org/}.
\end{itemize}

Although we separate here the different methods for combing trees, from now on AdaBoost, XGBoost and Random Forest will be referred to as BDTs.

\subsection{Cross-validation}

Cross-validation is a technique that allows us to assess how a classification generalizes to an independent data set. The idea behind cross-validation is to partition the data into independent subsets, training with one set while evaluating with the other. This can be done several times (number of folds) which allows one to train and measure the success rate of the classifier with the different samples ensuring that one is using information that is available in the entire simulation. In this work, we choose to do a 3-fold cross-validation which is enough to avoid over-fitting and to have a robust assessment of our model validity while maintaining a large training sample.

\subsection{Evaluating the classifier}\label{section:metrics}
The classification results in each SN being classified as a SNIa with the correct redshift or as a SNCC or SNIa with an inaccurate redshift. SNIa in the first group constitute the ``signal'' and those in the second group the ``background''.

The performance of a classifier can be evaluated using different metrics introduced in this section (e.g.~AUC). Since our goal is to obtain a large and reliable SNIa sample, it is natural to use purity and efficiency as indicators. The latter will also allow us to set a criterion for choosing a probability threshold. For efficiency and purity studies we will use a subset of the simulation to train and another independent subset to classify and therefore estimate the performance of the classifier.

\subsubsection{ROC curve}
Our problem is a binary one: events are either signal or background and are classified in these categories. A metric that is commonly used as an evaluation method for dichotomic classifications, is the AUC metric. AUC stands for Area Under Curve, where the curve is the ROC curve (Receiver Operating Characteristic). The ROC curve illustrates the performance of a binary classifier by plotting the true positive rate (efficiency) against the false positive rate (contamination). 

While the ROC curve represents the performance of a model in two-dimensions, the AUC simplifies this into a number. A perfect model would score an AUC of 1 while a random classification would score $0.5$.

\subsubsection{Purity and efficiency}\label{Section:pur_and_eff}
For a classified sample, purity and efficiency can be used as metrics. This requires a choice of the BDT response threshold.

Our goal in the classification of SN-like events is to obtain a sample of type Ia SNe with correct redshifts separated from other types of SNe or type Ia SNe with inaccurate redshifts. Redshifts are considered accurate if, when compared to the generated redshift for simulations, they satisfy $|\Delta z|/(1+z) < 0.1$, inaccurate redshifts exceed $0.15$. SNe Ia with inaccurate redshifts are not suitable for cosmological analysis. 

We define our efficiency or true positive rate as:
\begin{equation}
\epsilon_{Ia}=\frac{n^{true}_{Ia}}{N^{total}_{Ia}} \label{eq:true_positive}
\end{equation}
where $n^{true}_{Ia}$ are type Ia supernovae that have the correct redshift and are correctly classified and $N^{total}_{Ia}$ contain all synthetic SNIa before classification. In order to use the statistics of different cross-validation folds, we define the total efficiency as the weighted sum of each fold efficiency.

We make a distinction between: {\it total efficiency}, which is includes the detection, selection and classification steps; the {\it SN-like efficiency} which evaluates only the effect of selection cuts and classification; and the {\it classification efficiency}, which assesses our machine learning classification methods only. 

The purity of the SNIa sample is defined as:
\begin{equation}
P_{Ia}=\frac{n^{true}_{Ia}}{n^{true}_{Ia}+ n^{false}_{Ia}}
\end{equation}\label{eq:purity}
where $n^{true}_{Ia}$ is defined above and $n^{false}_{Ia}$ are either core-collapse that were classified as type Ia SNe, or type Ia supernovae with inaccurate redshifts.

Contamination by core collapse or SNe Ia with inaccurate redshifts is defined as:
\begin{equation}
C_i=\frac{n^{false \;i}_{Ia}}{n^{true}_{Ia}+ n^{false}_{Ia}} \label{eq:false_positive}
\end{equation}
where $i$ is the contaminating type.

Since our Ia and CC light-curve simulations are volumetric ones, SN rates are accounted for by weighting events when computing efficiencies and purity. We take volumetric rates for type Ia and core-collapse SNe from \citep{Pritchet:2008np} and \citep{Bazin:2009mp}, respectively. 

\subsection{Parameter setting and feature selection}

Learning algorithms have a set of chosen parameters, referred to as hyperparameters. In this work we implemented an automated search of hyperparameters using a grid of possible parameter values evaluated on our cross-validated sample with an appropriate score for each classifier (e.g. RF: mean accuracy, XGB: log loss). Using available scikit-learn tools to select hyperparameters and rank feature importance \citep{scikit-learn}.

Inefficient features were identified by measuring the impact of each feature in the model score. They were discarded in our classification. To avoid biasing this selection we used an iterative procedure where the order of evaluation of each feature was permuted at each round. 

For each classification method a different ranking of features was obtained. However, some features were selected as efficient for all methods. These were: photometric SN redshift, the $\chi^2_\nu$ of the agreement in color and width of the redshift fit, the rise time for $i$, $r$ and $g$ filters, the fall time for $i$ and $r$ filters, the $\chi^2_\nu$ of the general light-curve fit for $i$ and $r$ filters and the amplitude in $g$ for the general light-curve fitter.

\section{Comparison of different classification methods}

\subsection{Application on simulated SNe}
\begin{figure}[h!]
\begin{center}
\includegraphics[width=0.8\linewidth]{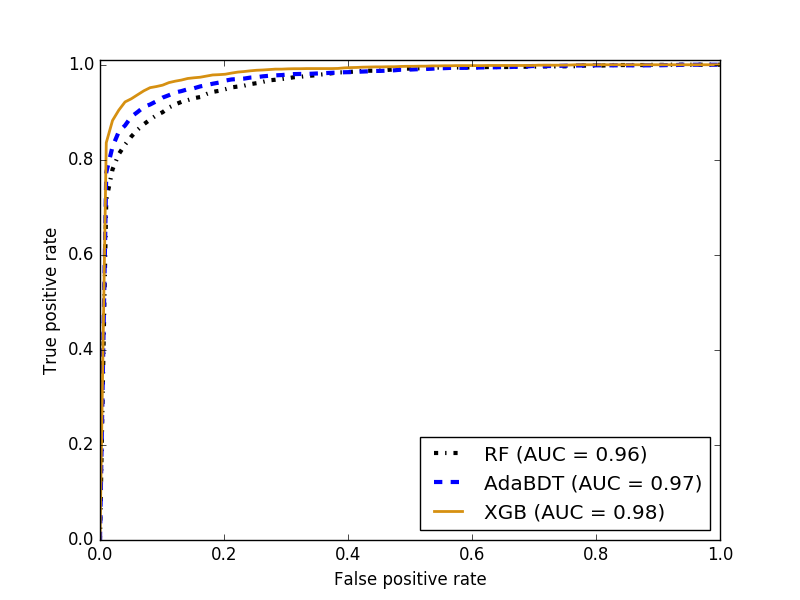}
\caption{Receiver operating characteristic (ROC) curves for different classification methods applied on synthetic SNe. True and false positive rates are given by Equations \ref{eq:true_positive} and \ref{eq:false_positive} respectively. Each curve represents a different algorithm: Random Forest RF (dotted black), AdaBoost Decision Tree AdaBDT (dashed blue) and Extreme Gradient Boosting XGB (solid yellow). The AUC score (area under the curve) is shown in the legend.}\label{fig:ROC}
\end{center}
\end{figure}

Three classification methods are examined: Random Forest, AdaBoost and XGBoost Decision Trees. We evaluate the performance of the different methods using the AUC metric from the ROC curve shown in Figure \ref{fig:ROC}. All methods obtain an AUC metric above $0.95$ and are therefore considered as excellent classifiers. This curve shows the trade-off between having a high positive rate (equivalent to classification efficiency) and a low false positive rate (equivalent to contamination). In the following, we study, the differences between methods and their impact on both simulated and real data.

\begin{figure}[h!]
\begin{center}
\includegraphics[width=0.8\linewidth]{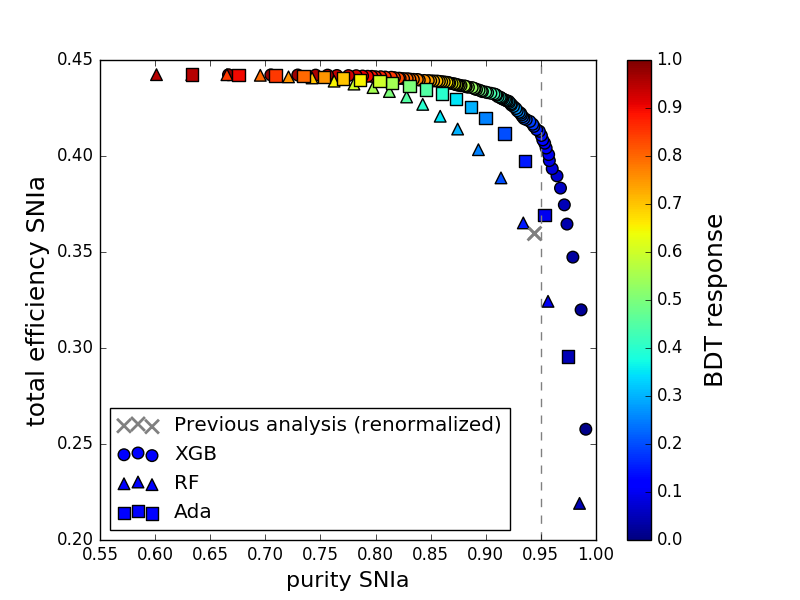}
\caption{Total efficiency (Equation \ref{eq:true_positive}) versus purity (Equation \ref{eq:purity}) for different choices of the BDT response cut for the 3 classification methods. The plot has been constructed for the simulated sample. The results for the XGB, RF and Ada classifiers are shown as circles, triangles, and squares respectively. The gray dashed line indicates a purity of $95\%$. The gray cross shows the performance of previous SNLS photometric classification (using host-galaxy redshifts \citep{Bazin:2011em}) with the efficiency renormalized upwards to account for the effect of redshift assignment ($83\%$).}\label{fig:eff_pur_NN}
\end{center}
\end{figure}

We classify synthetic SNe to estimate efficiencies and purities. For each classifier, a BDT response threshold must be chosen. This choice results from a trade-off between purity and efficiency of the classified sample. For our three methods, we plot in Figure \ref{fig:eff_pur_NN} total efficiency (as defined in Section \ref{Section:pur_and_eff}) against purity of the classified sample for different BDT response thresholds. The performance of each algorithm is in agreement with the AUC metric ranking. It is clear that the trade-off between efficiency and purity is more favorable for the XGB algorithm. 

To compare the three algorithms, we set the BDT response threshold such that the estimated purity is $95\%$. Total efficiencies and purities for these samples can be seen in Table \ref{table:global_eff_pur}. The algorithm with highest efficiency for our set purity is found to be XGB.

\subsubsection{Efficiency evolution with redshift}
\begin{figure}[h!]
\begin{center}
\includegraphics[width=0.8\linewidth]{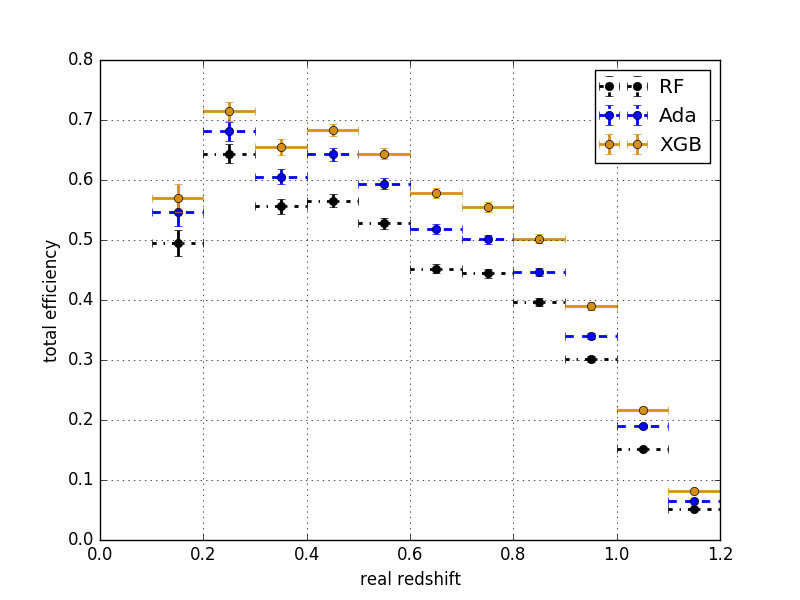}
\caption{Total efficiency from synthetic SNIa light-curves as a function of the real simulated redshift for different classification methods with the purity set to ~$95\%$. Random Forest (RF) points are plotted with black dotted error bars, AdaBoost Decision Tree (Ada) with dashed blue and Extreme Gradient Boosting (XGB) with solid yellow. Note that the total efficiency of XGB is higher than the other two methods at all redshifts.}\label{fig:superimposed_efficiencies}
\end{center}
\end{figure}

The total efficiency as a function of redshift is shown in Figure \ref{fig:superimposed_efficiencies} for all classification methods. The higher efficiency at low redshift can be attributed to higher quality light-curves for nearby SNe Ia. The SNIa classification efficiency varies from one algorithm to the other, XGB being the best performing method over the whole redshift range. Interestingly AdaBoost and XGB differences are quite homogeneous which can be attributed to the similarities of their optimization methods.

\subsubsection{Evolution of purity with redshift}

Figure \ref{fig:pur_z} shows the evolution of purity and contamination as a function of real and photometric SN redshifts for each algorithm. The contamination by type CC SNe is higher at lower real redshift but remains small (below $15\%$) whatever the method. Comparing the contaminations as a function of real redshift and photometric SN redshift, there is a migration of low-$z$ events towards higher redshifts. This is attributed to inaccurate photometric SN redshifts for some core-collapse events, as illustrated in Figure \ref{fig:sim_SN_zpho_gz}.

\afterpage{%
\begin{figure}
    \centering
    \begin{subfigure}[b]{0.45\textwidth}
        \includegraphics[width=\textwidth]{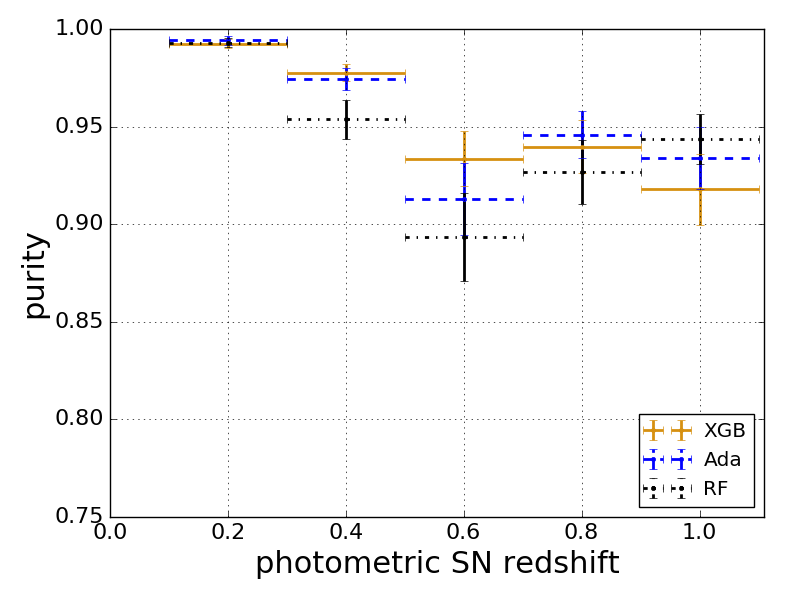}
        \caption{\footnotesize SNIa purity}
        \label{fig:pur_zpho_Ia}
    \end{subfigure}  
    \begin{subfigure}[b]{0.45\textwidth}
        \includegraphics[width=\textwidth]{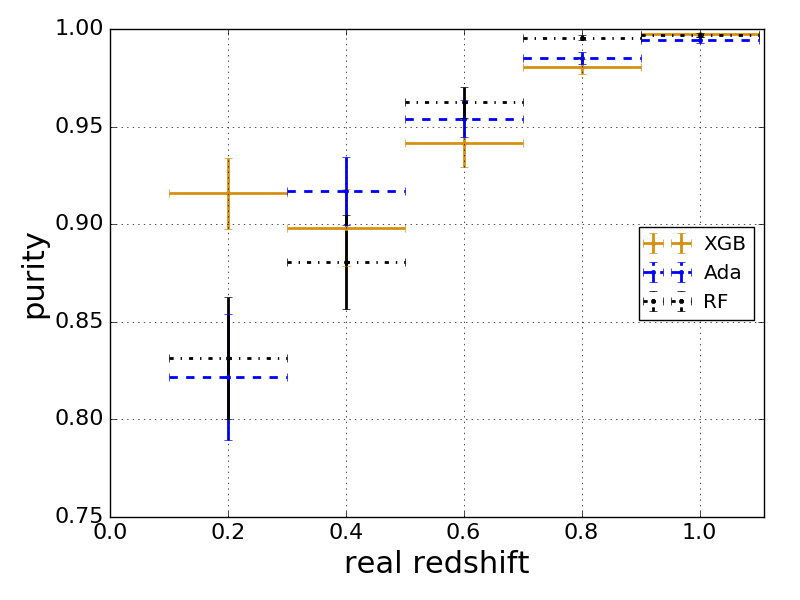}
        \caption{\footnotesize SNIa purity}
        \label{fig:pur_gz_Ia}
    \end{subfigure}
     
 	\begin{subfigure}[b]{0.45\textwidth}
        \includegraphics[width=\textwidth]{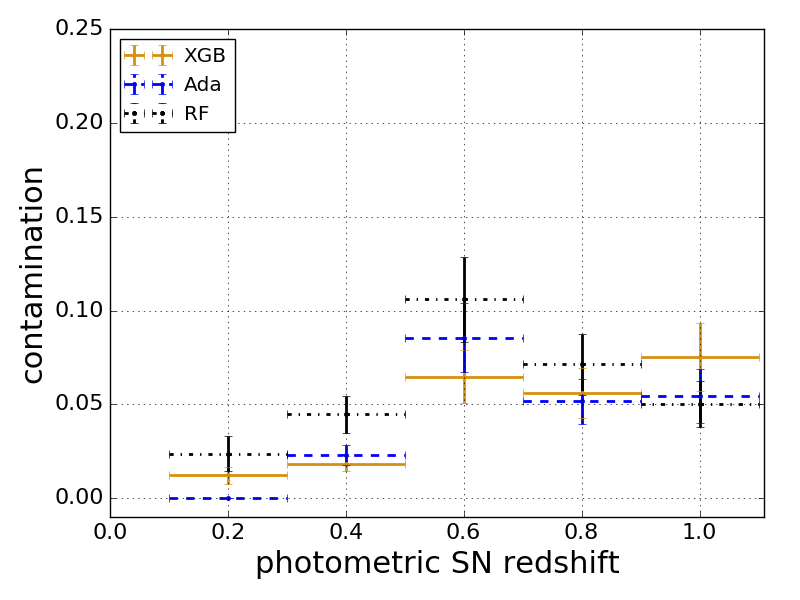}
        \caption{\footnotesize CC contamination}
        \label{fig:pur_zpho_CC}
    \end{subfigure}   
    \begin{subfigure}[b]{0.45\textwidth}
        \includegraphics[width=\textwidth]{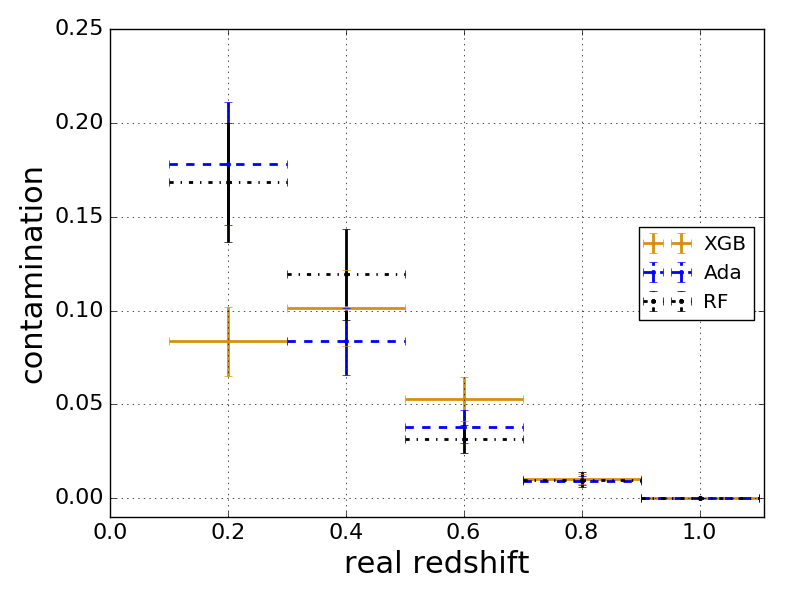}
        \caption{\footnotesize CC contamination}
        \label{fig:pur_gz_CC}
    \end{subfigure}
    \begin{subfigure}[b]{0.45\textwidth}
        \includegraphics[width=\textwidth]{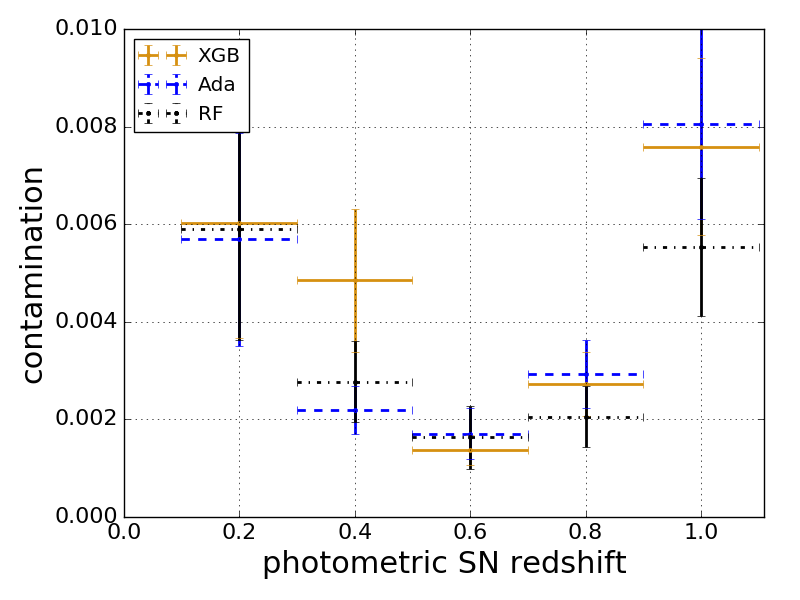}
        \caption{\footnotesize SNIa inaccurate z contamination}
        \label{fig:pur_zpho_Ia_badz}
    \end{subfigure}
    \begin{subfigure}[b]{0.45\textwidth}
        \includegraphics[width=\textwidth]{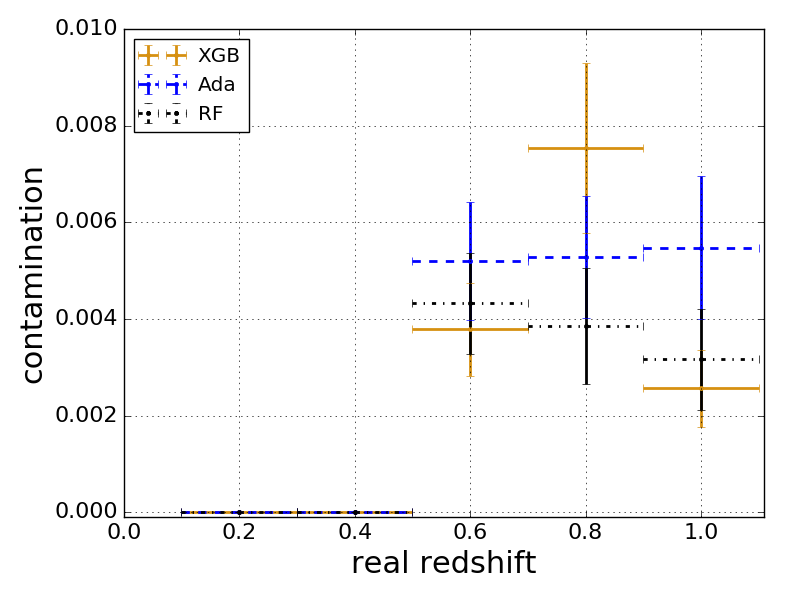}
        \caption{\footnotesize SNIa inaccurate z contamination}
        \label{fig:pur_gz_Ia_badz}
    \end{subfigure}
    \caption{ SNIa purity and background contamination from simulated SN light-curves as a function of redshift for different methods with a given purity of $95\%$. Left column : photometric SN redshift, right column : real redshift. Same color code as in Figure \ref{fig:superimposed_efficiencies}. Note how the CC SNe that are incorrectly classified as SNe Ia tend to have an assigned photometric SN redshift that is much higher than their real redshift.}\label{fig:pur_z}
\end{figure}
\clearpage
}

The contamination by type Ia SNe with inaccurate photometric SN redshift increases with higher redshift, but the overall contamination stays well below $1\%$.

\subsection{Application on SNLS3 data}
\begin{figure}[h!]
\begin{center}
\includegraphics[width=0.6\linewidth]{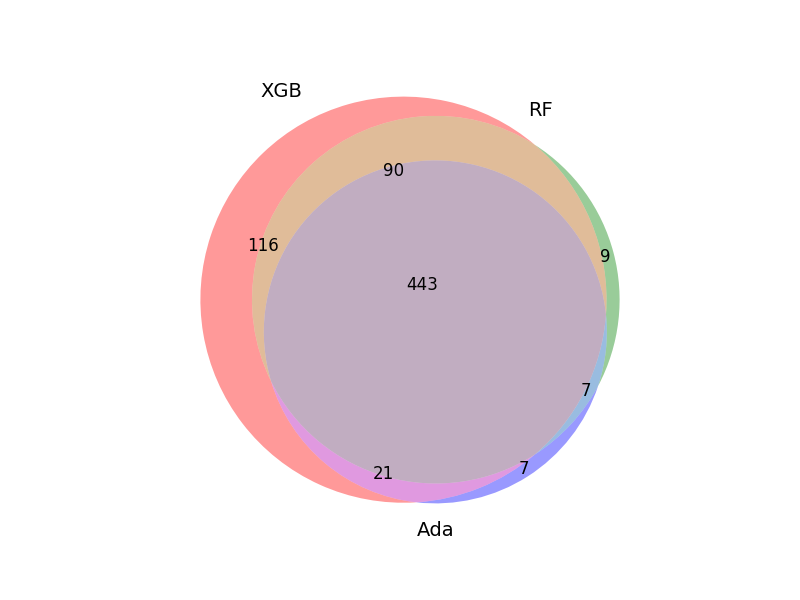}
\caption{Venn diagram for SNLS3 data photometrically classified by three different algorithms with the purity set to $95\%$. Color code as follows : overlap between the three methods in mauve, overlap between RF and XGB (Ada) in beige (light blue), between XGB and Ada in purple, pure RF in green, XGB in pink and Ada in dark blue. The total number of classified events for each method are given in Table \ref{table:photo_sample}.}\label{fig:venn_methods}
\end{center}
\end{figure}

Classification is also evaluated on SNLS3 selected data with the purity set to $95\%$. In Figure \ref{fig:venn_methods} we show a Venn diagram with the events classified as type Ia SNe by each method. The large number of common events shows the coherence between the three algorithms. 

\begin{figure}
    \centering
    \begin{subfigure}[b]{0.5\textwidth}
        \includegraphics[width=\textwidth]{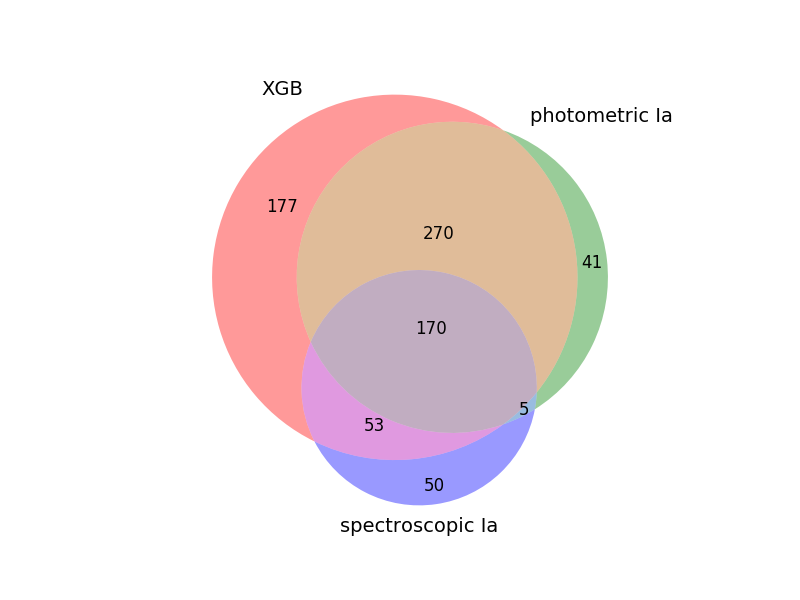}
        \caption{\footnotesize XGB}
        \label{fig:pur_zpho_Ia}
    \end{subfigure}  
    \begin{subfigure}[b]{0.5\textwidth}
        \includegraphics[width=\textwidth]{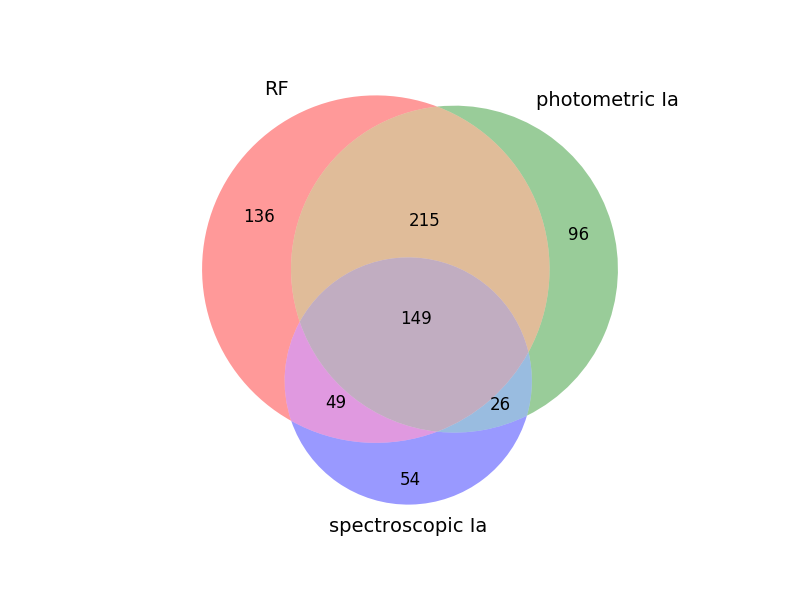}
        \caption{\footnotesize Random Forest }
        \label{fig:pur_gz_Ia}
    \end{subfigure}     
 	\begin{subfigure}[b]{0.5\textwidth}
        \includegraphics[width=\textwidth]{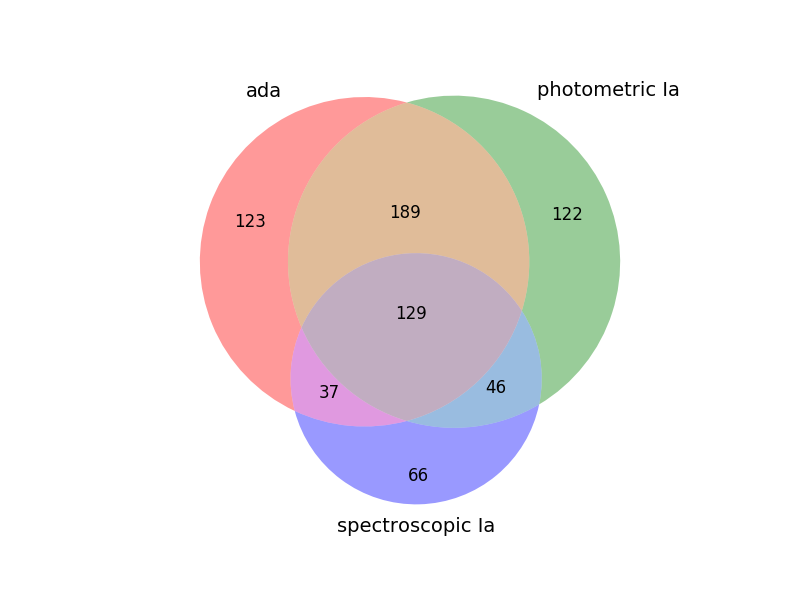}
        \caption{\footnotesize AdaBoost}
        \label{fig:pur_zpho_CC}
    \end{subfigure}
    \caption{ Venn diagrams for each classification algorithm, showing the intersection between events classified by our methods and the photometric and spectroscopic classifications. Dark blue and green regions correspond to events in those samples missed by our classification. }\label{fig:venn_methods_Ia}
\end{figure}

For each method, Table \ref{table:photo_sample} gives the number of events classified as type Ia for the selected sample and the sub-samples of photometrically and spectroscopically identified type Ia and core-collapse SNe. XGB has the largest number of spectroscopically and photometrically classified type Ia supernova. However, since these two samples share common events we visualize the superposition of these samples in Figure \ref{fig:venn_methods_Ia}. XGB continues to have the largest common sample.

For RF, all classified core-collapse events had an inaccurate photometric SN redshift. For AdaBoost, one event was found to have inaccurate photometric SN redshift. The other event, common to both spectroscopic and photometric samples, is a spectroscopically classified type II event whose light-curve is incomplete because it was observed at the end of a season. For XGB classification 4 core-collapse (2 spectroscopic and 2 photometric) events had an inaccurate photometric SN redshift. Four events had correct photometric SN redshifts (1 spectroscopic, 2 photometric and one common to both samples). One last event was classified photometrically, therefore no spectroscopic redshift was available. 

The XGB method selects more CC events than the other two algorithms. Given the expected CC contaminations (see Table \ref{table:global_eff_pur}) and photometric sample sizes (see Table \ref{table:photo_sample}), we expect XGB to classify $20$ to $40\%$ more CC events than the other two methods, less than what we observe in data on the two test-samples of CC events that we have at our disposal. This might be a statistical fluctuation, or a reflection of the incompleteness of the CC test-samples or an indication that our photometric samples are still contaminated by residual non SN-backgrounds that make our expected CC contaminations only indicative.

All classified samples contain the same spectroscopically confirmed SNIa with inaccurate photometric SN redshift when compared to its spectroscopic redshift. Using Tables \ref{table:global_eff_pur} and \ref{table:photo_sample}, the total number of type Ia SNe with inaccurate redshifts is expected to be between $1.5$ and $4$ events depending on the algorithm. This is in reasonable agreement with what we see in data on the sub-sample of spectroscopically identified SNe Ia for which we have both redshifts.

To check the agreement between expectations and data with the three methods, we compare the SNIa efficiency ratio between any two methods with the ratio of the classified sample sizes for the same two methods. The expected and observed ratios based on XGB and RF compare well. The two ratios defined with respect to AdaBoost are found to be higher in data than expected. This discrepancy remains unexplained. It may indicate that the XGB and RF samples are contaminated, in the same way, by non-SN backgrounds.

\begin{table}
\centering
\begin{tabular}{|c  c | c| c |c |}
	\hline
	 & & AdaBoost & Random Forest & XGBoost\\
	\hline
	total efficiency& Ia &$36.9\pm 0.6$&  $32.4\pm 0.7$ &$41.1\pm 0.7$\\
	purity & Ia & $95.6\pm 0.5$ & $95.6 \pm 0.4$& $95.3 \pm 0.4$\\
	contamination& Ia inaccurate z  & $0.53\pm0.09$& $0.29 \pm 0.07$ & $0.60 \pm 0.09$ \\
	contamination& CC &$3.9\pm0.4$& $4.1 \pm 0.4$ & $4.0 \pm 0.4$ \\
  	\hline
\end{tabular}
\caption{Estimated total efficiency, purity and contamination from simulated SNe for different methods with a given purity of ~$95\%$.}
\label{table:global_eff_pur}
\end{table}

\begin{table}
\centering
\begin{tabular}{| c | c| c |c |}
	\hline
	 & AdaBoost & Random Forest & XGBoost\\
	\hline
	photometric sample & 478 & 549 & 670\\
	\hline
	spectroscopic Ia & 166 &198 & 223\\
	photometric Ia & 318 & 364& 444\\
	spectroscopic CC & 2 & 2 & 3\\
 	photometric CC & 1& 1& 6\\
  	\hline
\end{tabular}
\caption{Events classified as SNe Ia by the three methods with purity set at ~$95\%$. The first line gives the numbers found from SNLS3 and the next four lines give the numbers for the spectroscopically and photometrically identified subsamples \citep{Guy:2010bp}, \citep{Bazin:2011em}, \citep{Bazin:2009mp}.}
\label{table:photo_sample}
\end{table}

\subsubsection{Comparison with the SNLS3 subluminous and peculiar SNIa samples}\label{section:sub_pec}
The SN-like sample (the starting point of our classification) contains 11 photometrically identified subluminous events and 5 spectroscopically identified peculiar events. 

For all classification methods, the same 3 peculiar events are contained in our photometrically classified sample. None of them exhibit any sign of peculiarity in their light-curves. The super-Chandrasekhar type Ia and the 1991T-like object are not classified as type Ia SNe by any of our methods.

Subluminous supernovae are found in our classified samples. In the case of Random Forest and AdaBoost classifications, 4 events are in the classified sample while 8 are included in the XGB sample. Despite our methods not being trained for disentangling normal type Ia and subluminous SNe, our photometric classification appears to have some efficiency in detecting subluminous SNe Ia as well.

\subsubsection{Effect of spectroscopic confidence index}\label{section:CI_95}

In Section \ref{section:SNLS3_data} we split the spectroscopically confirmed type Ia SNe according to the confidence level of the spectroscopic identification. Table \ref{table:CI_95} shows the percentage of events correctly classified for each method and sub-class. All three methods in this work have a larger classification efficiency for SNe Ia with the highest confidence index.

\begin{table}
\centering
\begin{tabular}{| c | c| c |c |}
	\hline
	 & AdaBoost & Random Forest & XGBoost\\
	\hline
	$\%$ SNIa ($CI=4$ or $5$) & $74 \pm 3 $ & $87 \pm 3$ & $96 \pm 2$  \\
	\hline
	$\%$ SNIa* ($CI=3$) &$58 \pm 6$ & $73 \pm 6$ & $88 \pm 4$ \\
  	\hline
\end{tabular}
\caption{Classification efficiency for our three methods when compared with the confidence index of the spectroscopic classification in \citep{Howell}.}
\label{table:CI_95}
\end{table}

\section{Choosing a method: XGB with high purity $98\%$}

The best performing algorithm was found to be XGB with high achievable purity and efficiency. We chose to select a sample with a purity of $98.0 \pm 0.3\%$. The corresponding total efficiency is $34.7 \pm 0.7 \%$. We now study this sample in detail.

\subsection{Effect of selection cuts and classification}

The impact of the selection cuts and the classification is shown for data and synthetic SNe Ia in Table \ref{table:sel_cuts_data}. The selection cuts (defined in Section \ref{Section:selectioncuts}) are shown to reduce the spectroscopic and photometric type Ia subsamples by $3.3\%$ and $1\%$ respectively. The core-collapse SNe are mainly discarded through classification. The two core-collapse events remaining after classification have inaccurate photometric SN redshifts. The classified sample contains 6 subluminous and 3 peculiar type Ia SNe from samples introduced in Section \ref{section:sub_pec}, and a spectroscopic type Ia that is classified with an inaccurate photometric SN redshift.

\begin{table}
\centering
\begin{tabular}{|c | c | c | c |c| c |c|}
	\hline
	cut & SNLS3 events & \multicolumn{2}{|c|}{spectroscopic} & \multicolumn{2}{|c|}{photometric} & simulated \\
	 &in sample & Ia & CC & Ia & CC  & Ia$\%$ \\
	\hline
	SN-like & 1483 & 246 & 42 & 486 & 109 & 50\\
	selected & 1193 & 238 & 30 & 481 & 77 & 47\\
	classified & 529 & 205 & 1 & 374 & 1 & 35\\
	\hline
\end{tabular}
\caption{Effect of selection cuts and classification (XGB) using SNLS3 data and synthetic type Ia. The classifier threshold is adjusted so that the purity is $98\%$.}
\label{table:sel_cuts_data}
\end{table}

\subsection{Classification and photometric SN redshifts}

\begin{figure}[h!]
\centering
\begin{center}
\begin{subfigure}[b]{0.49\textwidth}
	\includegraphics[width=\linewidth]{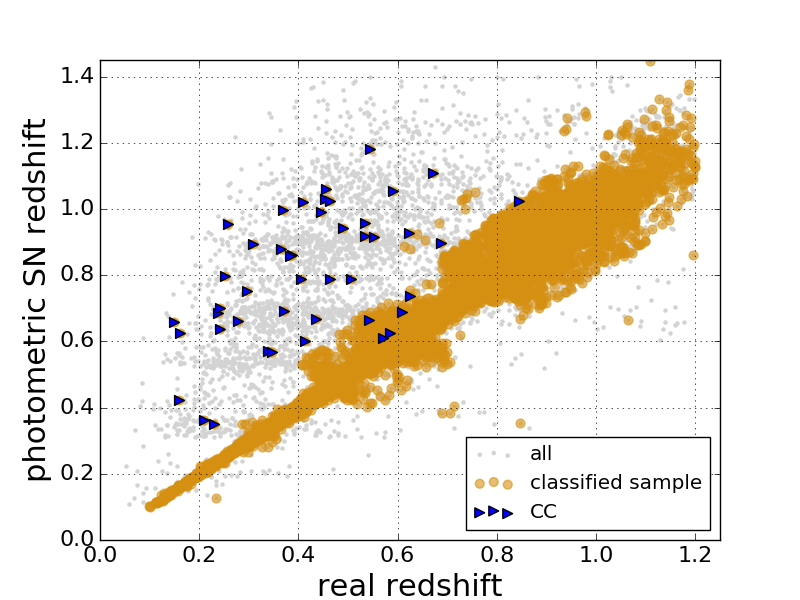}
	\caption{synthetic SNe in our photometric sample}\label{subfig:sim_SN_classified}
\end{subfigure}
\begin{subfigure}[b]{0.49\textwidth}
	\includegraphics[width=\linewidth]{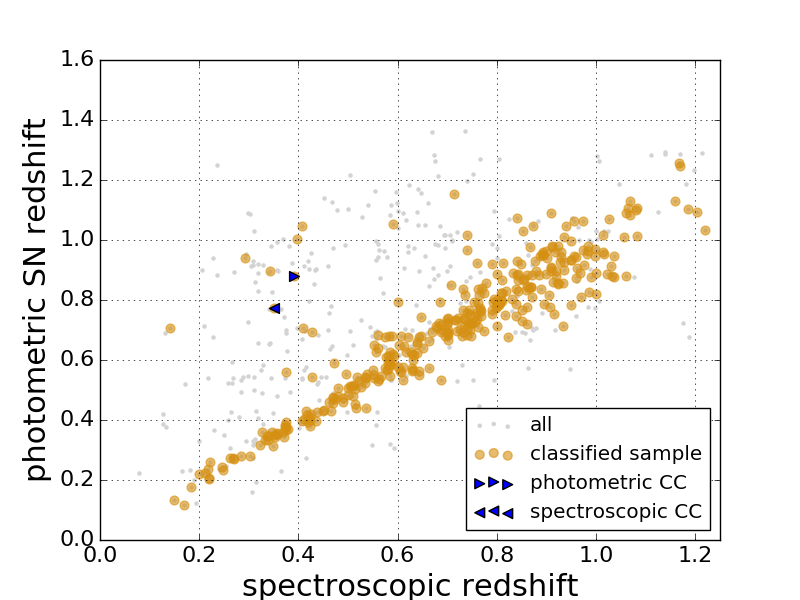}
	\caption{SNLS3 SNe in our photometric sample}\label{subfig:data_SN_classified}
\end{subfigure}
\caption{Photometric SN redshift versus real redshift for synthetic SNe (left) and SNLS3 data (right). All SNe in the SNLS3 selected sample are shown as gray dots. Events classified as type Ia with the XGB method at $98\%$ purity are shown as yellow circles. Blue triangles indicate core-collapse SNe classified as type Ia by our method.}\label{fig:gz_zpho}
\end{center}
\end{figure}

We investigate the impact the accuracy of the fitted photometric SN redshifts has on the classification. Figure \ref{fig:gz_zpho} shows the comparison between spectroscopic and photometric SN redshifts for events in the classified sample when both redshifts are available. Contamination by core-collapse SNe is mostly due to events that have an inaccurate photometric SN redshift. 

Interestingly, those core-collapse SNe that were assigned correct photometric SN redshifts were not classified as type Ia SNe by our method. A core-collapse event that has the correct photometric SN redshift is an event that has colors and photometry consistent with a type Ia supernova (photometric SN redshifts are obtained under the hypothesis that the object is a SNIa, see Section \ref{Section:SN_photo_z}). We highlight this rejection by our classification of core-collapse SNe with properties similar to SNe Ia and attribute it to the features obtained using the general SN fitter (see Section \ref{Section:feature_extraction}). 

\begin{figure}[h!]
\centering
\begin{center}
\begin{subfigure}[b]{0.48\textwidth}
\includegraphics[width=\linewidth]{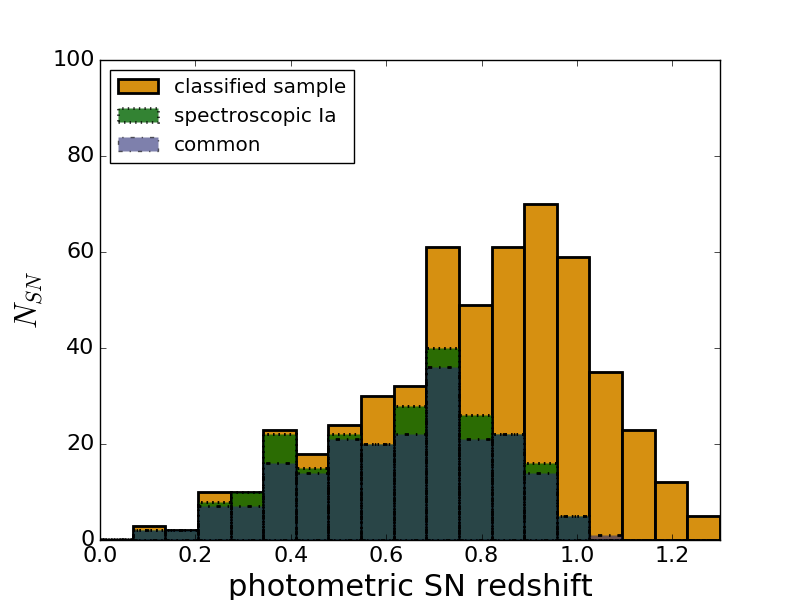}
	\caption{comparison with spectroscopic sample}\label{subfig:histo_new_spe}
\end{subfigure}
\begin{subfigure}[b]{0.48\textwidth}
\includegraphics[width=\linewidth]{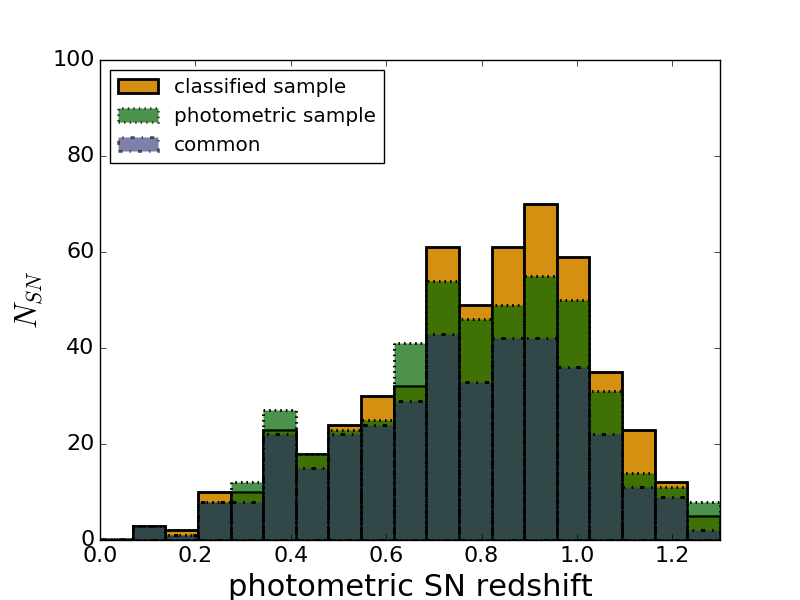}
	\caption{comparison with photometric sample}\label{subfig:histo_new_SNLS3}
\end{subfigure}
\caption{Photometric SN redshift distribution of the XGB classified SNLS3 sample in yellow (solid line) compared with spectroscopically (\ref{subfig:histo_new_spe}) and photometrically (\ref{subfig:histo_new_SNLS3}) identified test samples. Common events between samples are indicated. The photometric test sample is that obtained in \citep{Bazin:2011em} using host-galaxy redshifts.}\label{fig:histo_zpho}
\end{center}
\end{figure}

The photometric SN redshift distribution of classified events peaks at higher redshifts when compared to the spectroscopically identified sample (Figure \ref{subfig:histo_new_spe}). There is a large overlap between events in both samples and no particular trend over photometric SN redshift is seen.

The distribution of the SN-photometric redshift for the photometric sample classified in \citep{Bazin:2011em} and the one of this work are shown in Figure \ref{subfig:histo_new_SNLS3}. The new classification provides a larger number of $z>0.7$ events while maintaining the number of events at lower redshift, and therefore a large fraction of the spectroscopic and photometric  samples.

\subsection{Efficiency evolution: classification, selection and total}

\begin{figure}[h!]
\begin{center}
\includegraphics[width=0.8\linewidth]{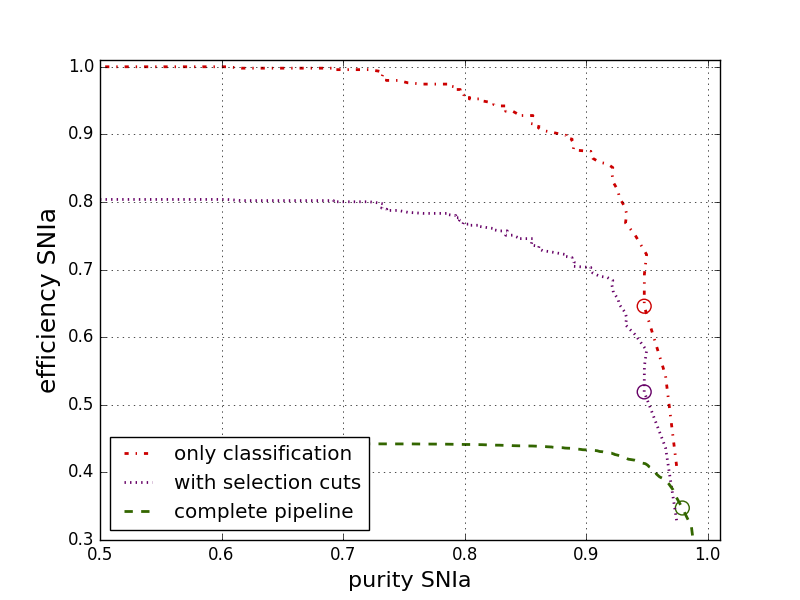}
\caption{Efficiency versus purity for XGB classification of simulated SNe. We show the trade-off between efficiency and purity when considering only classification effects (red dash-dot line), including selection cuts (purple dotted line) and taking into account the complete pipeline including detection (dashed green line). Circles indicate the evolution of efficiency and purity for the same BDT threshold (chosen such as to obtain a $98\%$ purity for the complete pipeline).}\label{fig:sup_eff_pur}
\end{center}
\end{figure}

In Figure \ref{fig:sup_eff_pur} we evaluate the efficiency-purity diagram taking into account: classification only, classification and selection cuts of Section \ref{Section:selectioncuts} and the complete pipeline. Our machine learning classification can achieve a $100\%$ efficiency at the expense of selecting a photometric sample with purity of at most $70\%$. The effect of adding selection cuts (necessary to ensure light-curve quality and reduction of non-SN backgrounds) is to reduce the maximum achievable efficiency to $80\%$. When accounting for the rest of the pipeline (detection and SN-like cuts) the maximum achievable efficiency drops to $45\%$. 

Despite a $\sim 20\%$ reduction in the maximum achievable efficiency between classification only and classification with selection cuts, we emphasize the importance of these cuts that reduce other possible non SN-like events. This is of great importance since our algorithms have been trained to disentangle type Ia from core-collapse SNe only. 

\subsection{The effect of the spectroscopic confidence index}\label{section:CI_98}
Spectroscopically identified type Ia SNe in SNLS3 with high confidence index $CI=4$ or $5$ (as defined in Section \ref{section:SNLS3_data}) have a photometric classification efficiency of $90 \pm 2 \%$. Those events with a $CI$ of $3$ have an efficiency in our analysis of $75\pm6 \%$. SNe with $CI=3$, are on average more distant than those with $CI=4$ and $CI=5$, so the photometry (and spectroscopy) will be noisier. This leads to a lower classification efficiency.

\subsection{The effect of light-curve quality}

\begin{figure}[h!]
\centering
\begin{center}
\begin{subfigure}[b]{0.48\textwidth}
	\includegraphics[width=\linewidth]{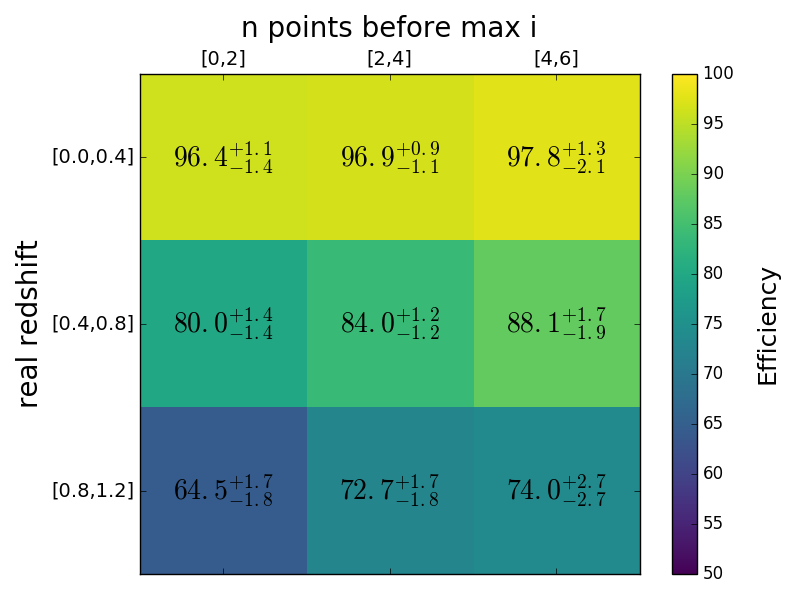}
	\caption{before maximum i filter}
\end{subfigure}
\begin{subfigure}[b]{0.48\textwidth}
	\includegraphics[width=\linewidth]{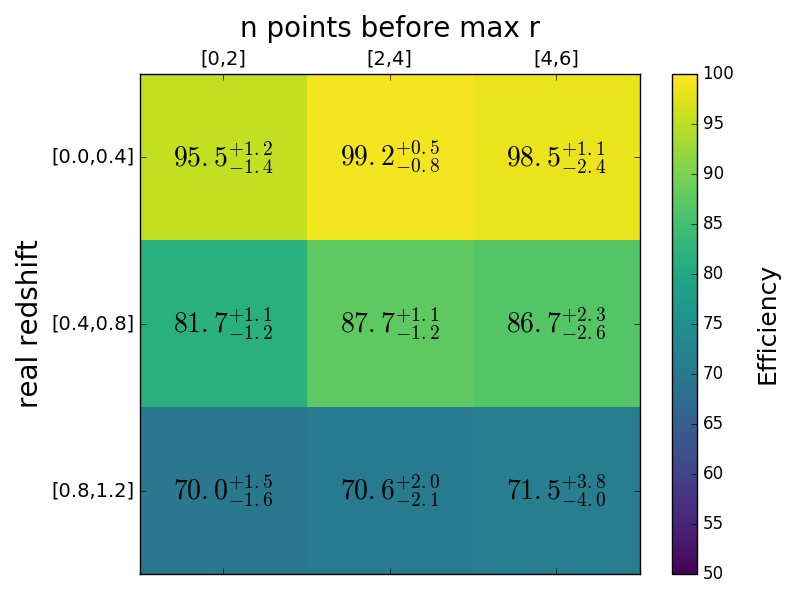}
	\caption{before maximum r filter}
\end{subfigure}
\begin{subfigure}[b]{0.48\textwidth}
	\includegraphics[width=\linewidth]{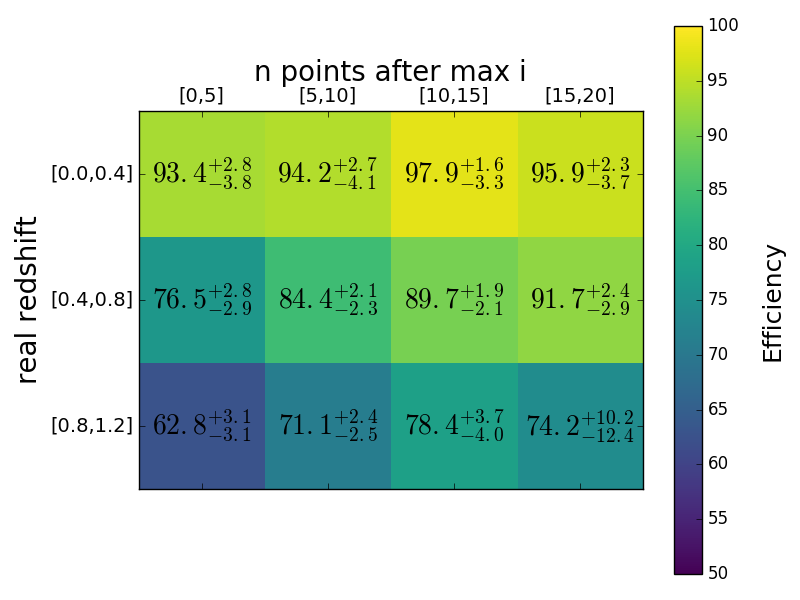}
	\caption{after maximum i filter}
\end{subfigure}
\begin{subfigure}[b]{0.48\textwidth}
	\includegraphics[width=\linewidth]{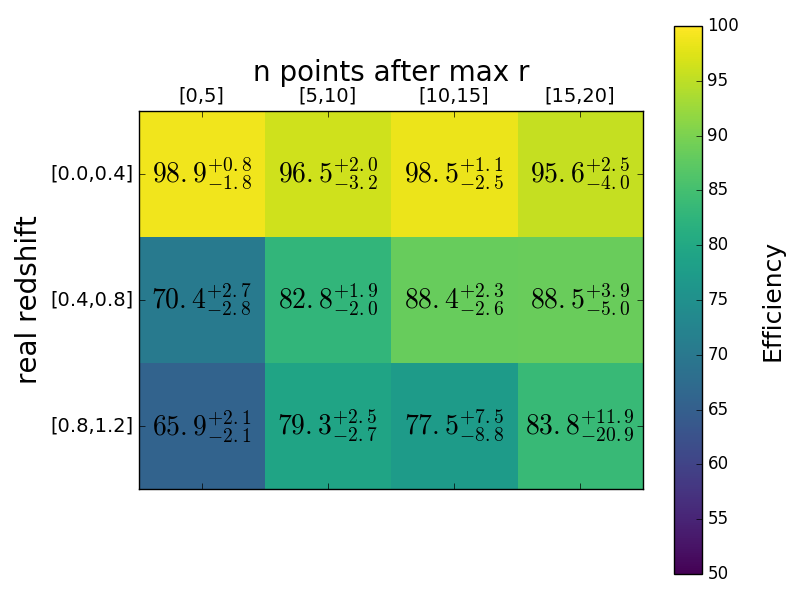}
	\caption{after maximum r filter}
\end{subfigure}
\caption{Percentage of simulated type Ia SNe correctly classified. The data is binned in redshift intervals (y axis) and the number of exposures before and after (top and bottom) maximum (left: i filter, right: r filter). The text indicates the percentage of correctly classified events in each bin with $1\sigma$ confidence intervals.}\label{fig:quality_Ia}
\end{center}
\end{figure}

We studied the performance of our classifier according to the quality of the available light-curves for type Ia SNe. The quality of light-curves was assessed through the number of exposures in the $i$ and $r$ filters before and after maximum light. In Figure \ref{fig:quality_Ia}, we show the percentage of correctly classified type Ia SNe as a function of the number of exposures in the previously mentioned filters.

For type Ia SNe, the larger the number of measurements after maximum, the higher the percentage of correctly classified events. This occurs for all redshift intervals. High redshift events require, as expected, better sampling to be correctly classified.

\section{Conclusions}

In this paper, we presented a new method for photometrically classifying type Ia supernovae using photometric redshifts derived from SN light-curves and machine learning techniques. This work is the first time that machine learning has been used to classify high redshift supernovae from photometry alone. We show that a sample of SNIa can be photometrically classified with a purity that is greater than $95\%$. Compared to previous work using external host photometric redshifts and sequential cuts, we obtain a purer sample at an equivalent efficiency.

We studied three different supervised learning algorithms for classification: Random Forest, and Decision Trees boosted using AdaBoost and XGBoost algorithms. We compared the results of the three classifiers using both simulated and real SN data. For a purity of $95\%$, we find that total efficiencies can vary by $10\%$ from one algorithm to the other, which may be linked to their optimization procedures. The XGB algorithm has the best performance both in terms of the AUC score (with a score of 0.98, with 1 being the perfect score) and the estimated efficiency when compared to other methods. When applied to real data, we obtain photometrically classified samples that are double the size of the spectroscopically confirmed sample in SNLS3. The coherence between the three algorithms can be seen from the large number of common classified events.

The best performing classification algorithm was found to be XGBoost. When trained with our synthetic SNe, it is able to provide a sample of $98\%$ purity and satisfactory efficiency. Core-collapse contamination is shown to be dominated by events with inaccurate redshifts. Interestingly, core collapse events with correct redshifts are properly classified as background by our method. These events have colors and photometry consistent with type Ia SNe and therefore should be harder to disentangle. This highlights the performance of our classification using features from a general SN light-curve fitter and the XGB algorithm.

In a real SN survey, efficiency is affected by different stages of the pipeline. In other classification studies, efficiencies and purities are computed directly from generated SN light-curves without taking into account selection cuts. These cuts are fundamental for the selection of a sample where non-modeled backgrounds are limited. The impact of selection cuts will vary with the pipeline and must be studied case by case. 

We find that selection cuts are fundamental for supernova photometric classification with supervised learning. An algorithm is only as good as its training set. Therefore if other backgrounds are present in the sample, the algorithm will perform less well. We argue that if the goal is SN classification, a substantial study must be done to ensure that non-modeled backgrounds are strongly reduced, and the extracted features are meaningful.

We acknowledge that our study was limited by the number of simulated SNe. Although $20,000$ simulated core-collapse were generated, when applying selection cuts to obtain the SN-like sample (the starting point of our classification) only a small percentage of core-collapse pass them. We expect a low number of core-collapse events at this stage, but it would be advisable to have a larger number to be used as training and for estimating efficiency and purity. This paper is a first step towards classification of real SN data using supervised learning and we will address this limitation in future work.

This work demonstrates for the first time the feasibility of machine learning classification in a high-redshift SN survey with application to real SN data. We have successfully classified a high-purity type Ia photometric supernova sample in the SNLS survey. An analysis of the impact on cosmology coming from the use of supervised learning techniques to produce SN samples will be subject of a future work. Additionally, this classification will be applied to the SNLS 5-year photometric analysis that will be the subject of a forthcoming paper.

\acknowledgments
Part of this research was conducted by the Australian Research Council Centre of Excellence for All-sky Astrophysics (CAASTRO), through project number CE110001020. 

AM thanks B. Schmidt, F. Yuan and B. Tucker for useful discussions.

This work was done based on observations obtained with MegaPrime/MegaCam, a joint project of CFHT and CEA/IRFU, at the Canada-France-Hawaii Telescope (CFHT) which is operated by the National Research Council (NRC) of Canada, the Institut National des Science de l’Univers of the Centre National de la Recherche Scientifique (CNRS) of France, and the University of Hawaii. This work is based in part on data products produced at Terapix available at the Canadian Astronomy Data Centre as part of the Canada-France-Hawaii Telescope Legacy Survey, a collaborative project of NRC and CNRS.

\bibliographystyle{JHEP}
\bibliography{myBib}

\providecommand{\href}[2]{#2}\begingroup\raggedright\begin{thebibliography}{10}

\bibitem{Riess:1998uy}
{\bf Supernova Search Team} Collaboration, A.~G. Riess et~al., {\it
  {Observational evidence from supernovae for an accelerating universe and a
  cosmological constant}},  {\em Astron.J.} {\bf 116} (1998) 1009--1038,
  [\href{http://arxiv.org/abs/astro-ph/9805201}{{\tt astro-ph/9805201}}].

\bibitem{Perlmutter:1999tu}
{\bf Supernova Cosmology Project} Collaboration, S.~Perlmutter et~al., {\it
  {Measurements of Omega and Lambda from 42 high redshift supernovae}},  {\em
  Astrophys.J.} {\bf 517} (1999) 565--586,
  [\href{http://arxiv.org/abs/astro-ph/9812133}{{\tt astro-ph/9812133}}].

\bibitem{Betoule:2014ui}
{\bf SDSS Collaboration} Collaboration, M.~Betoule et~al., {\it {Improved
  cosmological constraints from a joint analysis of the SDSS-II and SNLS
  supernova samples}},  {\em Astron.Astrophys.} {\bf 568} (2014) A22,
  [\href{http://arxiv.org/abs/1401.4064}{{\tt arXiv:1401.4064}}].

\bibitem{Bernstein:2011zf}
J.~P. Bernstein et~al., {\it {Supernova Simulations and Strategies For the Dark
  Energy Survey}},  {\em Astrophys. J.} {\bf 753} (2012) 152,
  [\href{http://arxiv.org/abs/1111.1969}{{\tt arXiv:1111.1969}}].

\bibitem{Abell:2009aa}
{\bf LSST Science Collaborations, LSST Project} Collaboration, P.~A. Abell
  et~al., {\it {LSST Science Book, Version 2.0}},
  \href{http://arxiv.org/abs/0912.0201}{{\tt arXiv:0912.0201}}.

\bibitem{Bazin:2011em}
G.~Bazin, V.~Ruhlmann-Kleider, N.~Palanque-Delabrouille, J.~Rich, E.~Aubourg,
  et~al., {\it {Photometric selection of Type Ia supernovae in the Supernova
  Legacy Survey}},  {\em Astron.Astrophys.} {\bf 534} (2011) A43,
  [\href{http://arxiv.org/abs/1109.0948}{{\tt arXiv:1109.0948}}].

\bibitem{Bazin:2009mp}
{\bf SNLS} Collaboration, G.~Bazin et~al., {\it {The Core-collapse rate from
  the Supernova Legacy Survey}},  {\em Astron.Astrophys.} {\bf 499} (2009) 653,
  [\href{http://arxiv.org/abs/0904.1066}{{\tt arXiv:0904.1066}}].

\bibitem{Ilbert}
O.~Ilbert, S.~Arnouts, H.~McCracken, M.~Bolzonella, E.~Bertin, et~al., {\it
  {Accurate photometric redshifts for the cfht legacy survey calibrated using
  the vimos vlt deep survey}},  {\em Astron.Astrophys.} {\bf 457} (2006)
  841--856, [\href{http://arxiv.org/abs/astro-ph/0603217}{{\tt
  astro-ph/0603217}}].

\bibitem{PalanqueDelabrouille:2009ng}
N.~Palanque-Delabrouille, V.~Ruhlmann-Kleider, S.~Pascal, J.~Rich, J.~Guy,
  et~al., {\it {Photometric redshifts for supernovae Ia in the Supernova Legacy
  Survey}},  {\em Astron.Astrophys.} {\bf 514} (2010) A63,
  [\href{http://arxiv.org/abs/0911.1629}{{\tt arXiv:0911.1629}}].

\bibitem{Kessler:2010wk}
R.~Kessler, A.~Conley, S.~Jha, and S.~Kuhlmann, {\it {Supernova Photometric
  Classification Challenge}},  \href{http://arxiv.org/abs/1001.5210}{{\tt
  arXiv:1001.5210}}.

\bibitem{Kessler:2010qj}
R.~Kessler et~al., {\it {Results from the Supernova Photometric Classification
  Challenge}},  {\em Publ. Astron. Soc. Pac.} {\bf 122} (2010) 1415--1431,
  [\href{http://arxiv.org/abs/1008.1024}{{\tt arXiv:1008.1024}}].

\bibitem{Karpenka:2012pm}
N.~V. Karpenka, F.~Feroz, and M.~P. Hobson, {\it {A simple and robust method
  for automated photometric classification of supernovae using neural
  networks}},  {\em Mon. Not. Roy. Astron. Soc.} {\bf 429} (2013) 1278,
  [\href{http://arxiv.org/abs/1208.1264}{{\tt arXiv:1208.1264}}].

\bibitem{Ishida:2012cf}
E.~E.~O. Ishida and R.~S. de~Souza, {\it {Kernel PCA for type Ia supernovae
  photometric classification}},  {\em Mon. Not. Roy. Astron. Soc.} {\bf 430}
  (2013) 509, [\href{http://arxiv.org/abs/1201.6676}{{\tt arXiv:1201.6676}}].

\bibitem{Lochner:2016hbn}
M.~Lochner, J.~D. McEwen, H.~V. Peiris, O.~Lahav, and M.~K. Winter, {\it
  {Photometric Supernova Classification With Machine Learning}},
  \href{http://arxiv.org/abs/1603.0088}{{\tt arXiv:1603.0088}}.

\bibitem{Regnault2009}
N.~{Regnault}, A.~{Conley}, J.~{Guy}, M.~{Sullivan}, J.-C. {Cuillandre},
  P.~{Astier}, C.~{Balland}, S.~{Basa}, R.~G. {Carlberg}, D.~{Fouchez},
  D.~{Hardin}, I.~M. {Hook}, D.~A. {Howell}, R.~{Pain}, K.~{Perrett}, and C.~J.
  {Pritchet}, {\it {Photometric calibration of the Supernova Legacy Survey
  fields}},  {\em AAP} {\bf 506} (Nov., 2009) 999--1042,
  [\href{http://arxiv.org/abs/0908.3808}{{\tt arXiv:0908.3808}}].

\bibitem{Guy:2010bp}
{\bf SNLS Collaboration} Collaboration, J.~Guy et~al., {\it {The Supernova
  Legacy Survey 3-year sample: Type Ia Supernovae photometric distances and
  cosmological constraints}},  {\em Astron.Astrophys.} {\bf 523} (2010) A7,
  [\href{http://arxiv.org/abs/1010.4743}{{\tt arXiv:1010.4743}}].

\bibitem{Guy:2007dv}
{\bf SNLS Collaboration} Collaboration, J.~Guy et~al., {\it {SALT2: Using
  distant supernovae to improve the use of Type Ia supernovae as distance
  indicators}},  {\em Astron.Astrophys.} {\bf 466} (2007) 11--21,
  [\href{http://arxiv.org/abs/astro-ph/0701828}{{\tt astro-ph/0701828}}].

\bibitem{Perret2010}
K.~{Perrett}, D.~{Balam}, M.~{Sullivan}, C.~{Pritchet}, A.~{Conley},
  R.~{Carlberg}, P.~{Astier}, C.~{Balland}, S.~{Basa}, D.~{Fouchez}, J.~{Guy},
  D.~{Hardin}, I.~M. {Hook}, D.~A. {Howell}, R.~{Pain}, and N.~{Regnault}, {\it
  {Real-time Analysis and Selection Biases in the Supernova Legacy Survey}},
  {\em AJ} {\bf 140} (Aug., 2010) 518--532,
  [\href{http://arxiv.org/abs/1006.2254}{{\tt arXiv:1006.2254}}].

\bibitem{Howell}
D.~A. {Howell}, M.~{Sullivan}, K.~{Perrett}, T.~J. {Bronder}, I.~M. {Hook},
  P.~{Astier}, E.~{Aubourg}, D.~{Balam}, S.~{Basa}, R.~G. {Carlberg},
  S.~{Fabbro}, D.~{Fouchez}, J.~{Guy}, H.~{Lafoux}, J.~D. {Neill}, R.~{Pain},
  N.~{Palanque-Delabrouille}, C.~J. {Pritchet}, N.~{Regnault}, J.~{Rich},
  R.~{Taillet}, R.~{Knop}, R.~G. {McMahon}, S.~{Perlmutter}, and N.~A.
  {Walton}, {\it {Gemini Spectroscopy of Supernovae from the Supernova Legacy
  Survey: Improving High-Redshift Supernova Selection and Classification}},
  {\em APJ} {\bf 634} (Dec., 2005) 1190--1201,
  [\href{http://arxiv.org/abs/astro-ph/0509195}{{\tt astro-ph/0509195}}].

\bibitem{GonzalezGaitan:2010iw}
S.~Gonzalez-Gaitan et~al., {\it {Subluminous Type Ia Supernovae at High
  Redshift from the Supernova Legacy Survey}},  {\em Astrophys. J.} {\bf 727}
  (2011) 107, [\href{http://arxiv.org/abs/1011.4531}{{\tt arXiv:1011.4531}}].

\bibitem{Balland:2009ka}
{\bf SNLS} Collaboration, C.~Balland et~al., {\it {The ESO/VLT 3rd year Type Ia
  supernova data set from the Supernova Legacy Survey}},  {\em Astron.
  Astrophys.} {\bf 507} (2009) 85, [\href{http://arxiv.org/abs/0909.3316}{{\tt
  arXiv:0909.3316}}].

\bibitem{Bronder:2007hp}
{\bf SNLS} Collaboration, T.~J. Bronder et~al., {\it {SNLS Spectroscopy:
  Testing for Evolution in Type Ia Supernovae}},  {\em Astron. Astrophys.} {\bf
  477} (2008) 717, [\href{http://arxiv.org/abs/0709.0859}{{\tt
  arXiv:0709.0859}}].

\bibitem{Ellis:2007hx}
{\bf SNLS} Collaboration, R.~S. Ellis et~al., {\it {Verifying the Cosmological
  Utility of Type Ia Supernovae: Implications of a Dispersion in the
  Ultraviolet Spectra}},  {\em Astrophys. J.} {\bf 674} (2008) 51--69,
  [\href{http://arxiv.org/abs/0710.3896}{{\tt arXiv:0710.3896}}].

\bibitem{Howell:2006vn}
{\bf SNLS} Collaboration, D.~A. Howell et~al., {\it {The type Ia supernova
  SNLS-03D3bb from a super-Chandrasekhar-mass white dwarf star}},  {\em Nature}
  {\bf 443} (2006) 308, [\href{http://arxiv.org/abs/astro-ph/0609616}{{\tt
  astro-ph/0609616}}].

\bibitem{scikit-learn}
F.~Pedregosa, G.~Varoquaux, A.~Gramfort, V.~Michel, B.~Thirion, O.~Grisel,
  M.~Blondel, P.~Prettenhofer, R.~Weiss, V.~Dubourg, J.~Vanderplas, A.~Passos,
  D.~Cournapeau, M.~Brucher, M.~Perrot, and E.~Duchesnay, {\it Scikit-learn:
  Machine learning in {P}ython},  {\em Journal of Machine Learning Research}
  {\bf 12} (2011) 2825--2830.

\bibitem{xgb}
T.~{Chen} and C.~{Guestrin}, {\it {XGBoost: A Scalable Tree Boosting System}},
  {\em ArXiv e-prints} (Mar., 2016) [\href{http://arxiv.org/abs/1603.0275}{{\tt
  arXiv:1603.0275}}].

\bibitem{Pritchet:2008np}
{\bf SNLS} Collaboration, C.~J. Pritchet, D.~A. Howell, and M.~Sullivan, {\it
  {The Progenitors of Type Ia Supernovae}},  {\em Astrophys. J.} {\bf 683}
  (2008) L25, [\href{http://arxiv.org/abs/0806.3729}{{\tt arXiv:0806.3729}}].

\end{thebibliography}\endgroup

\end{document}